\documentclass[]{aastex631} 
\usepackage{amsmath}
\usepackage{amsfonts} 
\usepackage{nicefrac}  
\usepackage[utf8]{inputenc} 
\usepackage[T1]{fontenc}  
\usepackage{url} 
\usepackage{rotating}
\usepackage{booktabs}   
\usepackage{xcolor}
\usepackage{microtype}    
\usepackage{graphicx}
\usepackage{dsfont}

\definecolor{seagreen}{rgb}{0.190, 0.525, 0.361}
\definecolor{ceruleanCrayola}{rgb}{0.224, 0.663, 0.859}
\definecolor{unberableRed}{rgb}{0.8, 0.06, 0.03}
\definecolor{anthracite}{rgb}{0.271, 0.270, 0.318}
\definecolor{refereebrown}{rgb}{0.11, 0.11, 0.0}
\definecolor{operamauve}{rgb}{0.718 0.518 0.655}
\definecolor{ugly}{rgb}{1 0 0}

\newcommand{\REFEREE}[1]{{#1}}

\received{XXX}
\revised{YYY}
\accepted{\today}
\submitjournal{ApJ}
\shorttitle{Interpretable ML for finding IMBHs}
\shortauthors{Pasquato et al.}

\begin{document}
\title{Interpretable machine learning for finding intermediate-mass black holes}
\author[0000-0003-3784-5245]{Mario Pasquato}
\affil{Département de Physique, Université de Montréal, Montreal, Quebec H3T 1J4, Canada}
\affil{Physics and Astronomy Department Galileo Galilei, University of Padova, Vicolo dell’Osservatorio 3, I–35122, Padova, Italy}
\affil{Mila - Quebec Artificial Intelligence Institute, Montreal, Quebec, Canada}
\affil{Ciela, Computation and Astrophysical Data Analysis Institute, Montreal, Quebec, Canada}
\author[0000-0001-9511-4649]{Piero Trevisan}
\affil{Department  of  Physics,  Università di Roma Sapienza, Piazzale Aldo Moro 2, 00185 Rome, Italy}
\affil{INAF-Osservatorio Astronomico di Roma, via Frascati 33, 00040 Monte Porzio Catone, Italy}
\author[0000-0001-9688-3458]{Abbas Askar}
\affil{Nicolaus Copernicus Astronomical Center, Polish Academy of Sciences, ul. Bartycka 18, 00-716 Warsaw, Poland}
\author[0000-0002-4728-8473]{Pablo Lemos}
\affil{Département de Physique, Université de Montréal, Montreal, Quebec H3T 1J4, Canada}
\affil{Mila - Quebec Artificial Intelligence Institute, Montreal, Quebec, Canada}
\affil{Ciela, Computation and Astrophysical Data Analysis Institute, Montreal, Quebec, Canada}
\affil{Center for Computational Astrophysics, Flatiron Institute, NY, USA}
\author{Gaia Carenini}
\affil{Département d'Informatique, École normale supérieure - PSL Research University, Paris, France.}
\author[0000-0001-8799-2548]{Michela Mapelli}
\affil{Institut f\"ur Theoretische Astrophysik, Zentr\"um f\"ur Astronomie, Albert-Ueberle-Strasse 2, D-69120, Heidelberg, Germany}
\affil{Physics and Astronomy Department Galileo Galilei, University of Padova, Vicolo dell’Osservatorio 3, I–35122, Padova, Italy}
\affil{INFN - Padova, Via Marzolo 8, I–35131 Padova, Italy}
\affil{INAF - Osservatorio Astronomico di Padova, Vicolo dell’Osservatorio 5, I-35122 Padova, Italy}
\author[0000-0002-8669-5733]{Yashar Hezaveh}
\affil{Ciela, Computation and Astrophysical Data Analysis Institute, Montreal, Quebec, Canada}
\affil{Département de Physique, Université de Montréal, Montreal, Quebec H3T 1J4, Canada}
\affil{Center for Computational Astrophysics, Flatiron Institute, NY, USA}
\begin{abstract} 
\noindent Definitive evidence that globular clusters (GCs) host intermediate-mass black holes (IMBHs) is elusive. Machine learning (ML) models trained on GC simulations can in principle predict IMBH host candidates based on observable features. This approach has two limitations: first, an accurate ML model is expected to be a black box due to complexity; second, despite our efforts to realistically simulate GCs, the simulation physics or initial conditions may fail to fully reflect reality. Therefore our training data may be biased, leading to a failure in generalization on observational data.
Both the first issue -explainability/interpretability- and the second -out of distribution generalization and fairness- are active areas of research in ML. Here we employ techniques from these fields to address them: we use the anchors method to explain an XGBoost classifier; we also independently train a natively interpretable model using Certifiably Optimal RulE ListS (CORELS). The resulting model has a clear physical meaning, but loses some performance with respect to XGBoost. We evaluate potential candidates in real data based not only on classifier predictions but also on their similarity to the training data, measured by the likelihood of a kernel density estimation model. This measures the realism of our simulated data and mitigates the risk that our models may produce biased predictions by working in extrapolation.
We apply our classifiers to real GCs, obtaining a predicted classification, a measure of the confidence of the prediction, an out-of-distribution flag, a local rule explaining the prediction of XGBoost and a global rule from CORELS.
\end{abstract}

\section{Introduction}
The detection of quasars at high redshift \citep[see, e.g.,][]{2011Natur.474..616M, 2023ApJ...943...67S,2023arXiv230801230M} requires a mechanism for the rapid assembly of supermassive black holes. Intermediate-mass black holes (IMBHs) bridge the gap between stellar-mass remnants and supermassive black holes, potentially playing an important role as seeds for the latter \REFEREE{(see, e.g.,  \citealt{{2019PASA...36...27W}} and \citealt{2021NatRP...3..732V} for two recent reviews).} 
Searches for IMBHs in the present-day Universe are being actively carried out, with focus on high-density environments such as the Galactic center \citep[][]{2016ApJ...816L...7O, 2018MNRAS.480.4684B, 2019PASJ...71S..21T, 2019ApJ...871L...1T, 2020ApJ...890..167T, 2023ApJ...942...46K, 2023arXiv230304067T}, dwarf galaxies \citep[][]{2016ApJ...817...20M, 2018MNRAS.478.2576M}, and globular clusters \citep[GCs; see e.g.][]{2008MNRAS.389..379M, 2010MNRAS.406.1049C, 2012ApJ...750L..27S, 2022MNRAS.516.1788S, 2009Natur.460...73F, 2014Natur.514..202B, 2016MNRAS.460.2025K, 2017Natur.542..203K, 2018NatAs...2..656L}. Regarding Galactic GCs, the most comprehensive study providing direct IMBH mass upper limits based on radio observations is \cite{2018ApJ...862...16T}, in the following T18.


IMBHs may form via repeated black hole (BH) mergers \citep[][]{2002MNRAS.330..232C, gerosa2017,2018ApJ...856...92F,2019PhRvD.100d3027R,2019MNRAS.486.5008A,2021MNRAS.505..339M, 2021MNRAS.507.5132D}.
This channel is supported by observational evidence from gravitational waves \citep[][]{2016ApJ...818L..22A, 2016PhRvX...6d1015A, 2016PhRvL.116x1103A, 2016PhRvL.116f1102A, 2017PhRvL.118v1101A, 2020PhRvL.125j1102A, 2020ApJ...900L..13A} but there are other proposed mechanisms, such as population III stars \citep[][]{2001ApJ...551L..27M, 2016MNRAS.460.4122R,2021ApJ...910...30T,2023MNRAS.525.2891C,2023MNRAS.524..307S}, runaway stellar collisions \citep[][]{1998MNRAS.298...93B, 1999A&A...348..117P, 2004Natur.428..724P, 2015MNRAS.454.3150G, 2016MNRAS.459.3432M, 2017MNRAS.472.1677S, 2019MNRAS.487.2947D, 2021MNRAS.507.5132D,2022MNRAS.512..884R,2023MNRAS.526..429A,2023MNRAS.521.3553R}, or primordial formation \citep[][]{2008MNRAS.388.1426K, 2017CQGra..34c5006D}.

Lacking a smoking-gun, uncontroversial direct IMBH detection, indirect detection methods based on IMBH dynamical effects in GCs are useful for compiling a list of candidate hosts for follow-up.
The observable effects \citep[see][for a recent and quite comprehensive review of IMBH detection attempts]{2017IJMPD..2630021M} include the presence of a central cusp in surface density \citep[][]{1976ApJ...209..214B, 1976ApJ...208L..55N} and/or velocity dispersion \citep[][]{1989ApJ...347..251P, 2006ASPC..352..269N, 2008ApJ...676.1008N, 2011A&A...533A..36L, 2012A&A...542A.129L, 2013A&A...552A..49L, 2013A&A...554A..63F, 2013ApJ...769..107L, 2016IAUS..312..181L, 2016MmSAI..87..563L, 2016IAUS..312..189L, 2017IAUS..316..240L, 2017MNRAS.464.3090A}, mass-segregation reduction \citep[][]{2004ApJ...613.1143B, 2008ApJ...686..303G, 2009ApJ...699.1511P, 2010ApJ...713..194B, 2013ApJ...768...26U, 2016ApJ...823..135P, 2023arXiv230205542D}, anomalous accelerations from pulsar timing \citep[][]{0004-637X-795-2-116, 2017MNRAS.468.2114P, 2018MNRAS.473.4832G, 2017Natur.542..203K}, high velocity stars \citep[][]{1991ApJ...383..587M, 2011A&A...533A..36L, 2018arXiv180807878F}, and other clues \citep[][]{2007MNRAS.381..103M, 2008A&A...489.1079P, 2010A&A...512A..35P, 2014MNRAS.444...29L, 2016MNRAS.460.2542P, 2017MNRAS.464.3090A}.

\subsection{Applying machine learning (ML) models to indirect IMBH detection}
The current prevailing approach for indirect IMBH detection is comparing a measured physical quantity and its prediction according to a theoretical model of the specific phenomena affected by the presence of an IMBH. These handcrafted physical models clearly present the advantage of full interpretability. But it is (at least in principle) possible that more flexible data-driven models can significantly outperform them.
On the other hand, several data-driven approaches to IMBH detection based on machine learning (ML), while promising, have at least two clear drawbacks: first, the general expectation is that a \REFEREE{high-performance model} must be very complex, virtually becoming a black box\footnote{\REFEREE{The notion of a black box, together with its opposites white box or glass box, is now commonplace in ML literature \citep[see, e.g., ][]{garrett2022glass} and has been attested at least since \cite{ashby1956introduction}, though it is uncertain when exactly it was introduced. For the purposes of this paper, it is understood that \emph{Black box ML models are predictive formulas, either too complicated for human understanding, or based on proprietary code with purposefully hidden calculations} \citep[][]{rudin2022black}.}}, with few notable exceptions. Such a model could be a deep neural network or a decision tree ensemble. These kinds of models are hard to interpret due to their complexity, \REFEREE{irrespective of the soundness of the statistical foundations on which they are built. For instance proving that a neural network is a universal function approximator \citep[][]{HORNIK1989359} is scant consolation for the fact that humans can only make sense of the inner workings of a trained neural network model through laborious analysis that resembles experimental biology more than mathematics \citep[this reverse engineering work constitutes the newborn field of `mechanistic interpretability' see, e.g.,][]{olah2017feature, carter2019activation, 2023arXiv230105217N}.}
Second, the model would have to be trained on simulations. Even in the unlikely event that simulations were to correctly model most of the relevant physics, they are quite likely to be run from unrealistic initial conditions, because constraining the early stages of cluster evolution is essentially still an open problem \citep[see e.g.][]{2022MNRAS.510.2097T}.
The simulated data on which a data-driven model will be trained can therefore be biased by containing e.g. spurious associations between variables or by under- or over-representing instances with certain characteristics. In this paper, we address both issues.

\subsection{Explainability versus native interpretability}
\REFEREE{We build a baseline black-box model based on XGBoost \citep[][]{chen2016xgboost}. We treat XGBoost as a black box since it is impossible for a human within a reasonable time frame to make sense of the weighed contributions of thousands of decision trees. Details on XGBoost are given in 
\REFEREE{Section~\ref{sec:methods}}. We show that the black-box model} behavior can be explained locally using anchors \citep[][]{ribeiro2018anchors}. Anchors are logical rules that apply in the vicinity of a given instance, explaining the black box model locally -in the sense that they are faithful to the underlying model in a neighborhood of a selected instance. In concrete, anchors are combinations of few binary choices obtained by thresholding one or more features (for instance, relaxation time greater than two Gyr and half-mass radius smaller than three pc), which predict the behavior of the underlying classifier (predicting IMBH host versus non-host) on the majority of data points around the one for which we want an explanation. They have been shown empirically to be readily understandable by humans \citep[][]{ribeiro2018anchors}. Indeed, the anchors we find are often physically meaningful as we discuss below. 

Anchors are one of the many schemes recently introduced to explain black box ML models. Such explanations are necessarily unable to fully capture the behavior of the model they are meant to explain (for instance, anchors are local explanations, and differ based on the data point we seek an explanation for; moreover, sometimes no suitable anchor exists given our requirements). If this were not the case, we could ditch the original model and use the explanations directly for prediction. Following this line of argument, natively interpretable models, i.e. models that are simple enough to be understandable by humans directly, have the advantage of not needing a post-hoc explanation. \cite{rudin2019stop} for instance has argued that, with properly engineered features, supervised problems on tabular data can be tackled effectively by interpretable models without the need to rely on black boxes.
We thus train a fully interpretable model based on decision rules,  Certifiably Optimal RulE ListS \citep[CORELS; ][]{angelino2017learning}, to assess how it compares to XGBoost in terms of performance. The CORELS model finds rules that have a straightforward interpretation in terms of GC physics; moreover the rule list it finds bears some resemblance to the anchor explanations for XGBoost.

\subsection{Training on simulations, predicting on observations}
A systematic discussion of how out-of-distribution generalization may impact the application of ML methods in the context of astronomy is \cite{2020arXiv201200066A}, in which the authors try to measure the increase in generalization error due to training on cosmological simulations and predicting on actual observational data. Their discussion clearly applies to our case, but it focuses mostly on measuring a global distance between the training data and the data set on which the models are eventually deployed, evaluating the effects on overall generalization error. In this paper we will focus on deciding whether individual predictions should be trusted; thus we will be interested in measuring how far an individual data point corresponding to an actual, observed physical system is from the simulated training data.

\section{Data}
\subsection{Training, validation and test datasets}
To build a training and validation sample for our ML  models, and to test them, we used results from the MOCCA-Survey Database I simulations described in detail by \cite{2017MNRAS.464L..36A}. 
The MOCCA-Survey Database comprises nearly 2000 star cluster simulations with different initial parameters that were carried out using the {\sc MOCCA} code \citep{2013MNRAS.429.1221H,2013MNRAS.431.2184G}, which is based on the Monte Carlo algorithm \citep{1971Ap&SS..14..151H, 1982AcA....32...63S, 1986AcA....36...19S} for treating the long-term evolution of star clusters. The orbit-averaged Monte Carlo method combines a statistical approach for the treatment of distant two-body interactions that drive the dynamical evolution of a star cluster with the particle based approach of \textit{N}-body methods \citep{2001MNRAS.324..218G,joshi2001,2013MNRAS.429.1221H,2022ApJS..258...22R}. This approach allows for the implementation of important physical processes and the possibility to simulate star cluster models with several hundred thousands to millions of stars within a few days to weeks. MOCCA incorporates prescriptions for stellar and binary evolution based on the SSE/BSE codes \citep{2000MNRAS.315..543H, 2002MNRAS.329..897H}. For computing the outcome of close dynamical interactions between binary-single stars and binary-binary stars, {\sc MOCCA} uses the {\sc fewbody} code \citep{2004MNRAS.352....1F} which is a direct \textit{N}-body integrator for small-\textit{N} gravitational dynamics. MOCCA also implements a realistic treatment of escape processes in tidally limited clusters based on \citet{2000MNRAS.318..753F}.
In order to model the Galactic potential, MOCCA uses a simple point-mass approximation. The treatment of escapers from the tidally limited GC models is based on \citet{2000MNRAS.318..753F}. MOCCA has been comprehensively tested and compared with results from direct \textit{N}-body codes \citep[see e.g.][]{2013MNRAS.431.2184G, 2014MNRAS.445.3435H,2016MNRAS.458.1450W,2017MNRAS.470.1729M, 2019MNRAS.487.2412G}.

The star clusters models simulated in the MOCCA-Survey Database I span a wide range of initial parametes with different number of objects, metallicity, binary fraction and parameter distribution, central concentration, tidal and half-mass radii, and different natal kick prescriptions for stellar-mass BHs (see Table~1 in \citealt{2017MNRAS.464L..36A}).
Initial stellar masses in each cluster model were sampled using the \cite{2001MNRAS.322..231K} initial mass function (IMF) with minimum and maximum stellar masses of $0.08 \rm \ M_{\odot}$ and $100.0 \ M_{\odot}$. The initially densest ($\rho_{c} > 10^{6} \rm \ M_{\odot} \rm pc^{-3}$) GC models form an IMBH ($\ge 500 \ \rm M_{\odot}$) within a few hundred Myr of dynamical evolution through a combination of collisions, mergers and mass transfer events onto a seed BH. This seed typically forms from a merger between a BH and a very massive main-sequence star ($> 50 \ \rm M_{\odot}$). The latter typically form via the runaway merger of stars during the \REFEREE{early evolution (within 100 Myr) of these dense GC models. The rapid formation of a seed IMBH in these MOCCA models has been described as what is known as the \emph{fast} IMBH formation scenario \citep{2015MNRAS.454.3150G}.}

In other more moderately dense models ($10^{4} \rm \ M_{\odot} pc^{-3} \lesssim \rho_{c} \lesssim 10^{6} \rm \ M_{\odot} pc^{-3}$), the IMBH forms after a few Gyr of dynamical evolution from mergers during binary interactions involving a stellar-mass BH. \REFEREE{This has been referred to as the \emph{slow} IMBH formation scenario.}
Both the \emph{fast} and \emph{slow} IMBH formation mechanisms in the MOCCA GC models have been discussed in detail in \citet{2015MNRAS.454.3150G}. In most of these simulated cluster models, for mergers involving a BH and a star, it is assumed that the BH accretes the entirety of the mass of the star that it merges with \citep{2020MNRAS.498.4287H,2021MNRAS.502.2682A}.
Furthermore, gravitational wave recoil kick following the merger of two stars are also not implemented \citep{2018MNRAS.481.2168M,2022MNRAS.514.5879M}. Both these assumptions facilitate the formation and growth of IMBHs in these cluster models.

At any rate, any set of simulations having a shot at realism will explore a large space of parameters that may lead to potentially unphysical outcomes. For instance, stellar evolution parameters such as the amount of mass fallback in the event of supernova explosion may affect the formation of stellar-mass black holes, leading to the simulations spanning a wider range of mass-to-light ratios with respect to typical Galactic GCs. However our knowledge of real star clusters is not \REFEREE{complete}: recent evidence suggests that exploring a wide range of mass-to-light ratios \REFEREE{(e.g. 0.8 to 5.03 in our simulated clusters)} may indeed be a feature rather than a bug, given the possibility of e.g. hypercompact star clusters \citep[][]{2021ApJ...917...17G}. The important safeguard is that the potential lack of realism of the simulations used to train a ML model is taken into account and mitigated appropriately, as we discuss below.

\begin{figure}
\centering
    \includegraphics[width=0.9\linewidth]{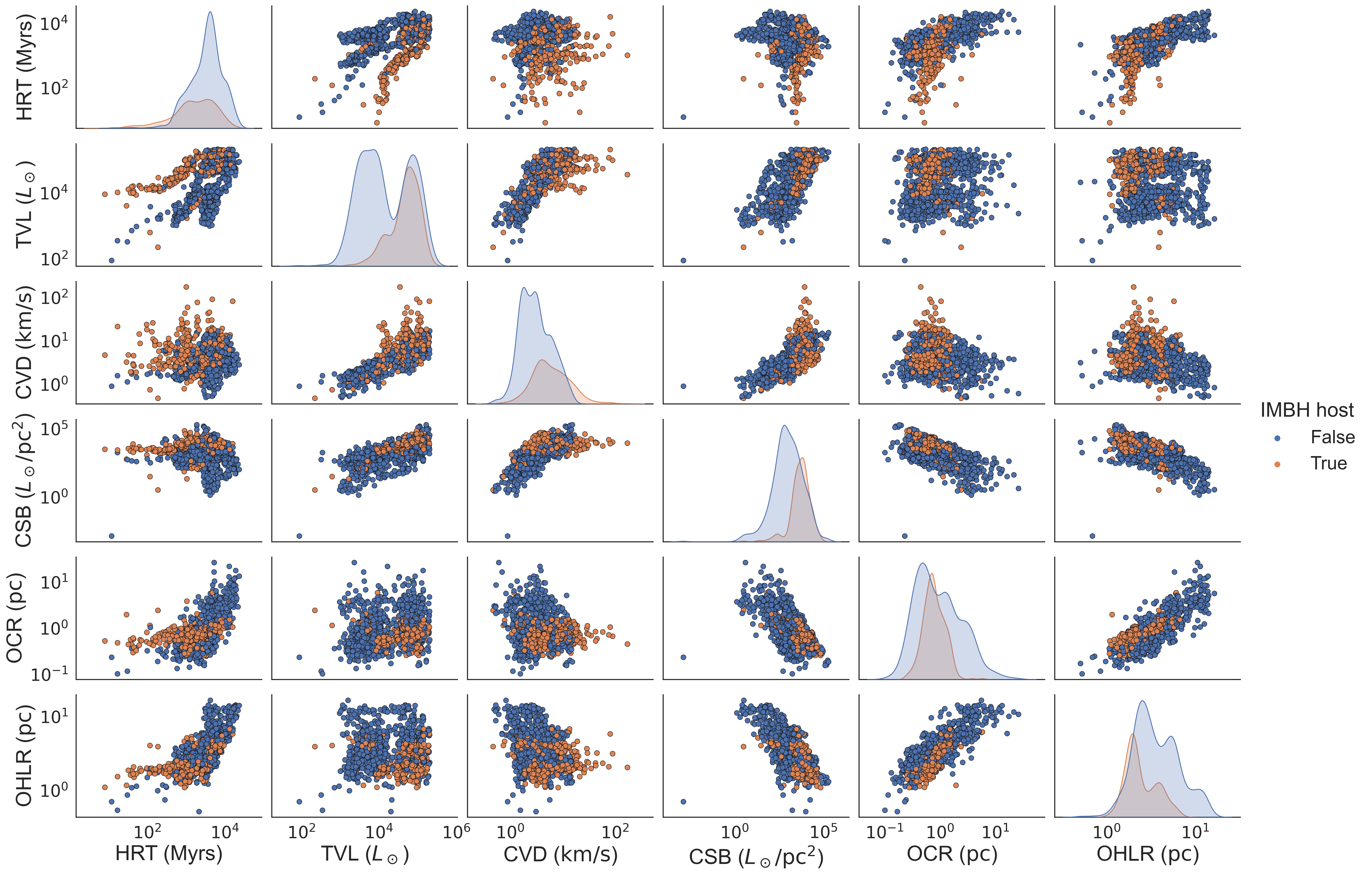}
\caption{Pair plot showing our GC simulations in feature space. IMBH hosts are shown as orange points and non-hosts as blue points. In order, the axis are: Half-mass relaxation time (HRT), total luminosity (TVL),   central velocity dispersion (CVD),  central surface brightness (CSB), core radius (OCR), and  half-light radius (OHLR). All the reported values are at $12$ Gyr. \label{fig:pairplot}}
\end{figure}

From the $1298$ GC models that evolved up to at least 12 Gyr in the MOCCA-Survey Database I, $388$ harboured an IMBH more massive than $500 \ M_{\odot}$; in the following we adopted a more inclusive definition of IMBH, extending the range to $100 \ M_{\odot}$ and above: this choice makes the classification problem harder and so is more conservative. Similar to the approach taken in \citet{2019MNRAS.485.5345A}, we used the simulation snapshot at 12 Gyr for each GC model to determine the following features: \cite{spitzer2014dynamical} half-mass relaxation time corresponding to the half-mass radius (HRT), total luminosity of cluster in units of solar luminosity (TVL), mass-weighted central velocity dispersion  (CVD) \REFEREE{in units of $\rm{km}/\rm{s}$},  central surface brightness in units of $\rm L_{\odot}/\mathrm{pc}^2$ from \cite{king1962structure} fitting of cumulative luminosity profile (CSB), core radius in $\mathrm{pc}$ obtained from \cite{king1962structure} (OCR), Half-light radius obtained from \cite{king1962structure} fitting in $\mathrm{pc}$ (OHLR). \REFEREE{Fig.~\ref{fig:pairplot} shows a pair plot of these features on the adopted dataset from MOCCA-Survey Database I.}

The simulation data was partitioned into a train-validation dataset, where we used k-fold cross validation as described below, and an internal test dataset comprising $10\%$ of the initial sample. This was stratified with respect to the label, i.e. each subset was forced to contain approximately the same proportion of IMBH hosts as the whole sample.

\subsection{Deployment dataset}
After training our models on simulated data, we used them to predict the presence of an IMBH on real GCs. To this end we use the observational features obtained by \cite{2018MNRAS.478.1520B} on a sample of $161$ GCs in the Milky Way as a final dataset to deploy our models. The data was obtained by accessing \url{https://people.smp.uq.edu.au/HolgerBaumgardt/globular/parameter.html}. \REFEREE{Observational data can in principle be out-of-distribution with respect to the simulations comprising our training set. We discuss how we are mitigating this issue by identifying out-of-distribution points using Kernel Density Estimation (KDE) in Section~\ref{KDE}.}

\section{Methods}
\label{sec:methods}

\subsection{K-fold cross-validation}
K-fold cross-validation is a widely used technique in ML to evaluate the generalization performance of a model on a given dataset. The process of k-fold cross-validation involves partitioning a dataset into k subsets, or folds, of approximately equal size. For each iteration, k-1 of the folds are used as training data, and the remaining fold is used as the validation set. This process is repeated k times, with each fold serving as the validation set once. The average performance of the model across all k iterations is then used as an estimate of its performance on unseen data. The major advantage of this technique is that it allows for a more robust estimate of the model's performance by utilizing all of the data for both training and validation, thus reducing the risk of overfitting. In this work,  we utilize $k=5$ and stratify the sampling so to have roughly the same fraction of IMBH hosts and non-hosts in each fold.

\subsection{Performance metrics}
\REFEREE{We evaluate the performance of our classifiers in terms of two metrics: precision and recall. Precision, also referred to as purity in astronomical catalogs, is the fraction of relevant instances (real IMBH hosts) among the retrieved instances (IMBH host claims), while recall, also known as completeness, is the fraction of the total amount of relevant instances (real IMBH hosts) that were actually retrieved (claimed to be hosts).}
\REFEREE{If a classifier returns a number between $0$ and $1$ as a measure of its confidence that a given instance belongs to the designated class ($0.642$ meaning for instance that a given GC is predicted to be an IMBH host with $64.2\%$ probability), it is possible to choose a threshold to convert such number to a hard prediction. For any given threshold we can thus obtain a value for precision and one for recall. Repeating this procedure for different thresholds we obtain a precision-recall curve, which illustrates the trade-off between precision and recall for different threshold values of a binary classifier. This curve shows the ability of our classifier to maintain a balance between avoiding false positives and capturing true positives as the classification threshold is varied.}
\REFEREE{If on the other hand a classifier returns only a hard classification, this procedure cannot be carried out and only one (precision, recall) couple is calculated.}

\REFEREE{It is worth pointing out that the nature of the problem of IMBH detection calls for a comparison mostly on the precision metric. Precision is more important than recall because false positives (IMBH claims in the absence of an actual IMBH) would waste telescope time and other resources needed for an IMBH-candidate follow-up. On the other hand, just one confirmed claim would have momentous astrophysical implications, so false negatives (missed IMBH hosts) are not problematic if at least some true positive is found.}

\subsection{Learning algorithms}
\subsubsection{XGBoost classifier}
We used XGBoost \citep[Extreme Gradient Boosting; ][]{chen2016xgboost}. \REFEREE{XGBoost found wide application in astronomy \citep[][]{2017AJ....153..204S, 2019MNRAS.489.4741S, 2021MNRAS.503.3279S}.} XGBoost is a scalable and efficient implementation of gradient boosting, which is an ensemble learning method that combines multiple weak decision trees to form a strong predictor. 
Decision trees are a supervised learning algorithm. They work by recursively splitting the dataset into smaller sub-groups based on the features that lead to the highest improvement in the prediction accuracy. The final result is a tree-like model where the internal nodes represent the decisions based on the input features, and the leaf nodes represent the prediction or outcome. 

The decision trees in XGBoost are grown using a gradient-based optimization algorithm that aims to minimize a cost function that measures the prediction error. The final prediction is obtained by combining the predictions of all trees through a weighted sum, where the weights are learned during training.

In comparison to random forests, which also use an ensemble of decision trees, XGBoost grows trees sequentially and tries to correct the mistakes of previous trees at each step. This results in a stronger predictor, but with a higher computational cost \citep[][]{chen2016xgboost}. 
\REFEREE{Humans can effectively track of the decisions of only a few shallow decision trees at most \citep[][]{freitas2014comprehensible}. Therefore,} all tree ensemble methods, unlike single trees, lack interpretability as soon as the number of trees grows enough to justify their use \citep[][]{molnar2022}, with the lone exception of the recently introduced FIGS algorithm \citep[][]{tan2022fast}. 

\begin{figure}[ht]
    \centering
    \begin{minipage}[b]{0.45\linewidth}
        \centering
        \includegraphics[width=\textwidth]{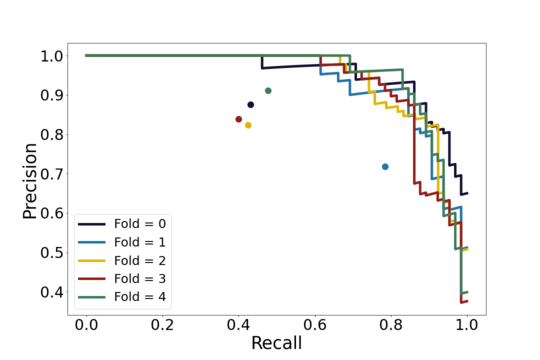}
    \end{minipage}
    \hspace{0.5cm}
    \begin{minipage}[b]{0.45\linewidth}
        \centering
        \includegraphics[width=\textwidth]{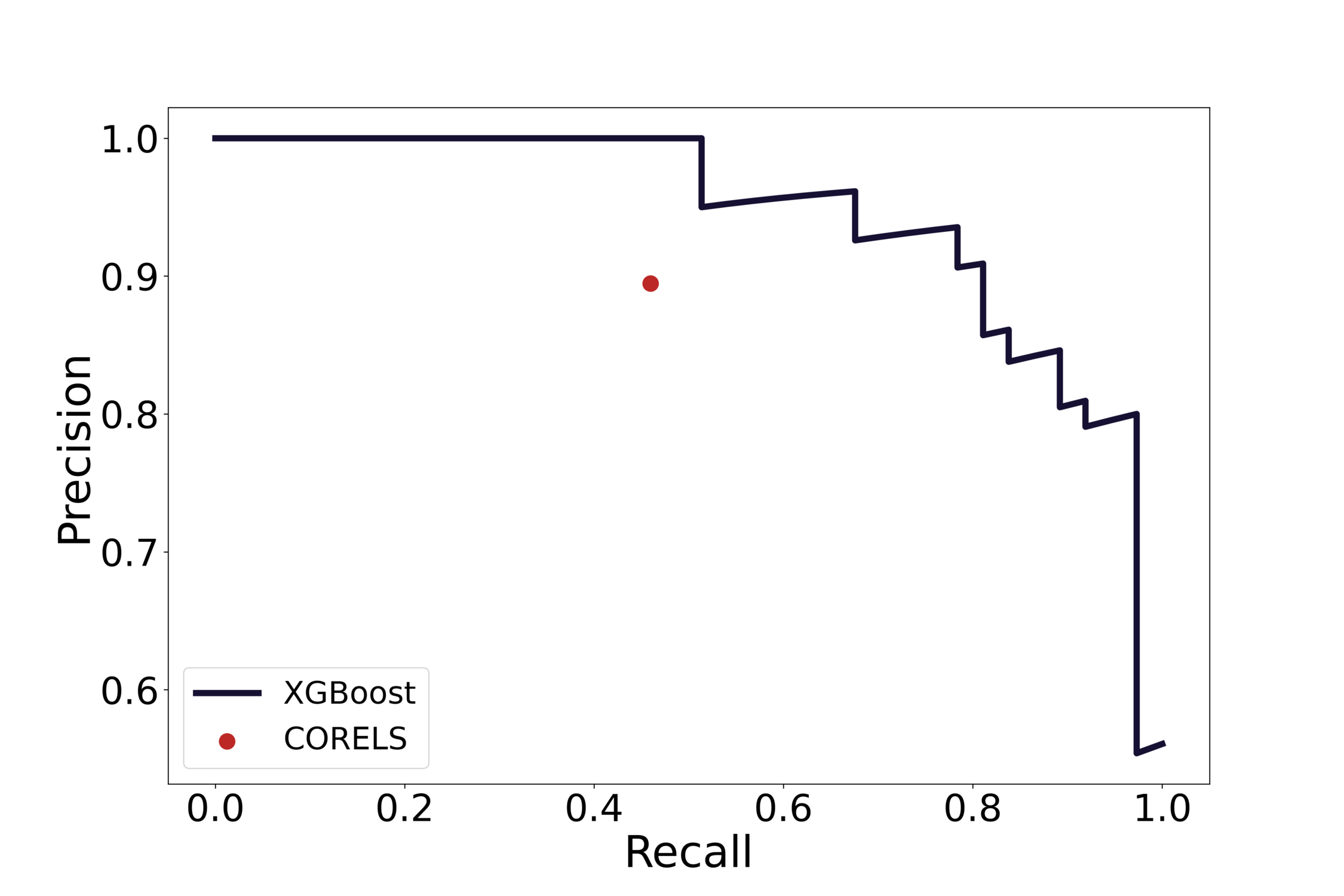}
    \end{minipage}
\caption{Left-hand panel: precision recall curve for our XGBoost classifier on each cross-validation fold (solid lines). The dots represent the recall and precision achieved by a CORELS classifier trained on the same fold. The colours identify the fold for both XGBoost and CORELS. Right-hand panel: Precision recall curve for our XGBoost classifier on the test set. The red dot represents the recall and precision achieved by the CORELS classifier on the same test set. \REFEREE{In both panels CORELS yields a single point rather than a curve because it produces a fixed rule list leading to a hard class label.}}
\label{fig:prc}
\end{figure}

We trained an XGBoost classifier based on an ensemble of $1000$ trees of maximum depth $2$ on our set of six features. 
We held out $20\%$ of our simulation data set for validation. Fig.~\ref{fig:prc} shows a precision-recall curve for our classifier on the test set. As discussed above, being based on a large ensemble of decision trees, our classifier is not readily interpretable: it is a black box. In Fig.~\ref{fig:prc} we compare its performance \REFEREE{in terms of its precision-recall curve} with that of a fully interpretable classifier trained on the same dataset, which we will discuss below.  \REFEREE{The left-hand side of Fig.~\ref{fig:prc} shows the precision-recall curves obtained in cross-validation, while the right hand side shows the final curve on the test set.} 

\subsubsection{Explaining XGBoost classifier: Anchors}
\cite{ribeiro2018anchors} introduced local, model agnostic explanation in the form of anchor rules as an improvement on their previous algorithm, LIME \citep[][]{2016arXiv160204938T}. Anchors are decision rules -sequences of if-then logical statements- that locally approximate a black box: one such rule list is learned for each data point for which an explanation is required. The scope of applicability of such a rule is explicitly defined in terms of the neighborhood of the relevant data point, leading to the definition of anchors as scoped rules. Over the data points to which the rule applies, i.e. those that satisfy the logical conditions, the black box prediction coincides with the prediction cast for the data point being explained, at least with a given frequency, termed precision. The fraction of points covered by the rule is termed coverage. \cite{ribeiro2018anchors} empirically show that anchors are readily understandable to humans when used as post-hoc explanations to a black box model. Anchors however have a few drawbacks: first, being a post-hoc explanation they capture the behavior of the underlying black box only locally and up to the chosen precision; second, they can get quite complex or have very low coverage, especially in the proximity of the decision boundary, which challenges either their usefulness or their effectiveness as human-understandable explanations. 

For instance, an application for a loan could be rejected by a black-box decision algorithm; the relevant explanation could be a rule such as "you carry a balance on two credit cards or more, and your FICO score is below \REFEREE{$X$}"; this may not apply to other individuals, but is guaranteed to explain the majority of the relevant classification outcomes in a neighborhood of the instance for which we requested an explanation. In the vicinity of the decision boundary, finding such rules can become challenging. The intuition for this is that the decision is harder in such cases, making it harder to summarize the black box behavior in simple terms.






\subsubsection{CORELS}
\cite{angelino2017learning} introduced the CORELS algorithm. CORELS produces concise rule lists based on binary features up to a user-specified depth. These optimal rules are found by optimizing a loss function on the set of all possible rule lists satisfying the given constraints. 
Unlike previously used methods based on greedy search \citep[e.g.][]{rivest1987learning}, the advantage of CORELS is that it finds the optimal rule list given the constraints. Finding optimal rule lists of this kind is in general a computationally hard problem. CORELS solves it by reducing the search space of rule lists relying on recent mathematical results \REFEREE{in discrete optimization, including those proved by the authors \citep[][]{angelino2017learning} and in \citep[][]{nijssen2010optimal}}. It also introduces specialized data structures that keep track of intermediate computations and exploits symmetries to make the problem tractable. 
CORELS loss function is made up of two terms, the first one counting misclassifications plus an additional term penalizing the size (i.e. number of rules) of the rule list \citep[see Eq.~1 in][]{rudin2019stop}. CORELS is still slower than traditional greedy approaches for learning decision trees, but it has been shown to be more effective than those on a variety of real-life datasets \citep[see e.g.][]{rudin2018learning}.

 The trade-off between rule accuracy and simplicity in CORELS can be adjusted by the user by changing a hyper-parameter $\lambda$ that determines the relative weight of the two terms of the loss. For instance, setting $\lambda = 0.01$ means that we would be indifferent to misclassifying $1\%$ of our dataset if this means obtaining a rule list that contains one less rule.

Figure~\ref{fig:prc} shows the precision-recall curve of XGBoost on each of the five cross validation folds. On each fold we also trained a CORELS model. Both the performance of XGBoost and CORELS change across folds, but in the case of CORELS it is immediately evident \emph{why}. The only fold in which CORELS learned a different rule from the others (corresponding to the blue dot in Fig.~\ref{fig:prc}) achieves a much greater recall at the cost of a lowered precision. We show below that the rule learned by CORELS in this case is less strict in weeding out GCs that have a longer relaxation time.  

It is also quite evident from Fig.~\ref{fig:prc} that the folds on which XGBoost performs best (green and black solid lines respectively), see also a higher performance by CORELS, which achieves the best (green dot) and second best (black dot) precision in each of them.

\begin{table}
\centering
\caption{ Rules found by CORELS during training in 5-fold cross-validation. The first column reports the number of folds in which a given rule was found. The corresponding rule is reported in the second column.  \label{rules_found_by_corels}}
\begin{tabular}{ll}

\hline \hline
Number of folds & Rule\\
\hline
4 & $T_r < 2000$~\REFEREE{Myr} and $R_{h} < 3.0$~\REFEREE{pc} and both $L > 10^4$~\REFEREE{$L_\odot$} and $R_c/R_h > 0.2$ \\
1 & $T_r < 5000$~\REFEREE{Myr} and CSB $ > 10^3$~\REFEREE{$L_{\odot}/\mathrm{pc}^2$} and $L > 10^4$~\REFEREE{$L_\odot$} and $R_c/R_h > 0.2$ \\
\hline
\end{tabular}

\end{table}

\begin{figure}
    \centering
    \begin{tabular}{cc}
    \includegraphics[width=0.45\textwidth]{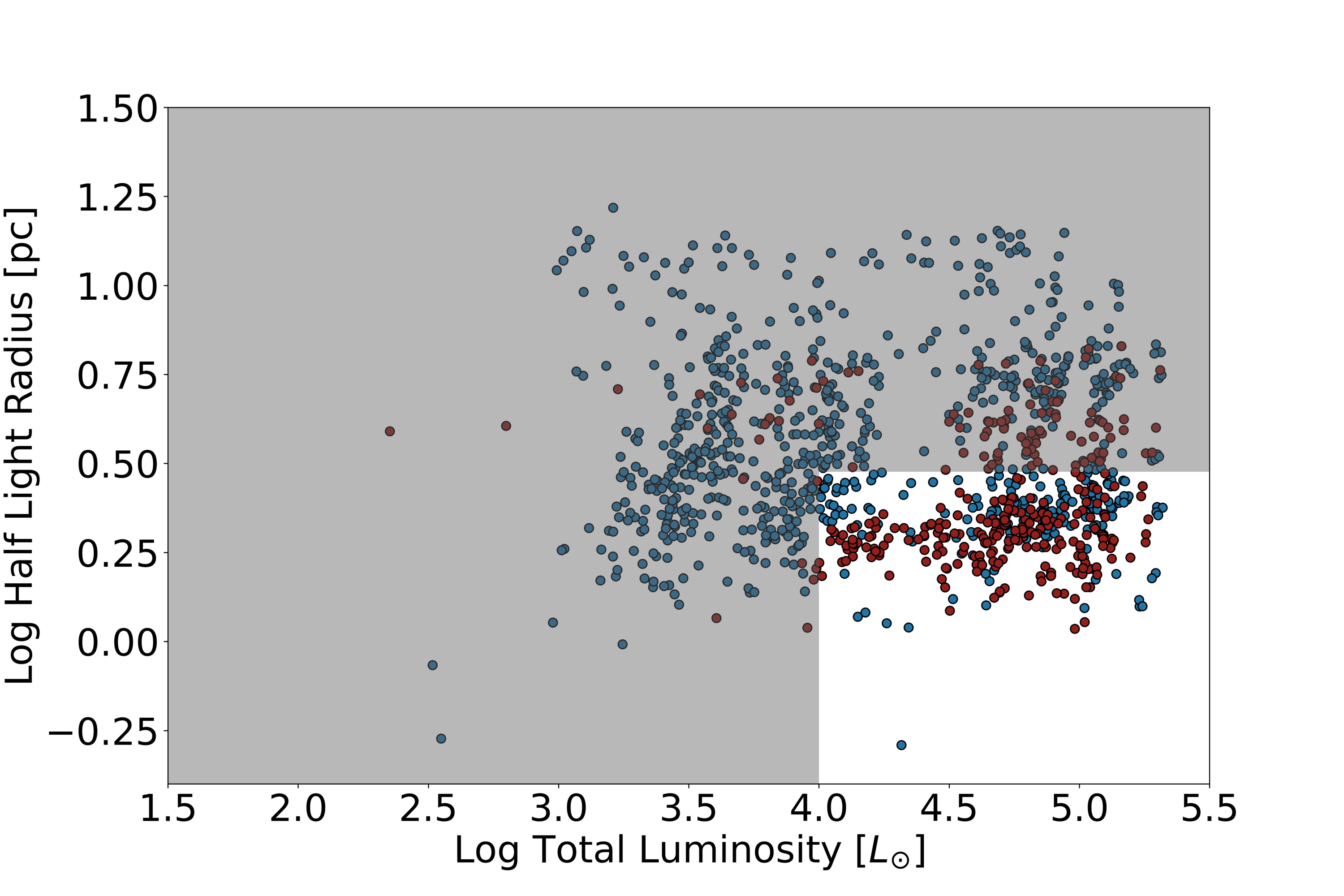} &  \includegraphics[width=0.45\textwidth]{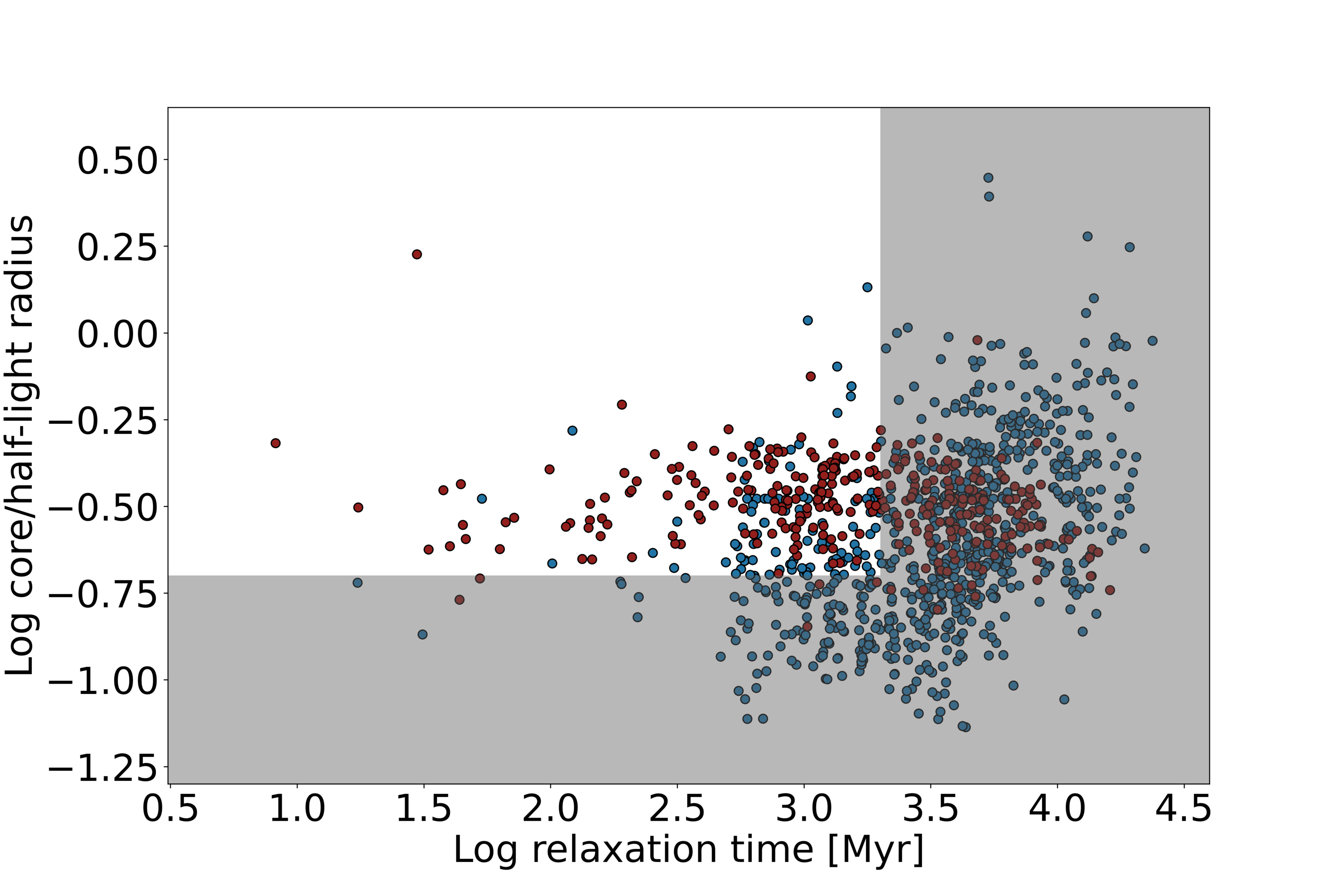} \\
    \includegraphics[width=0.45\textwidth]{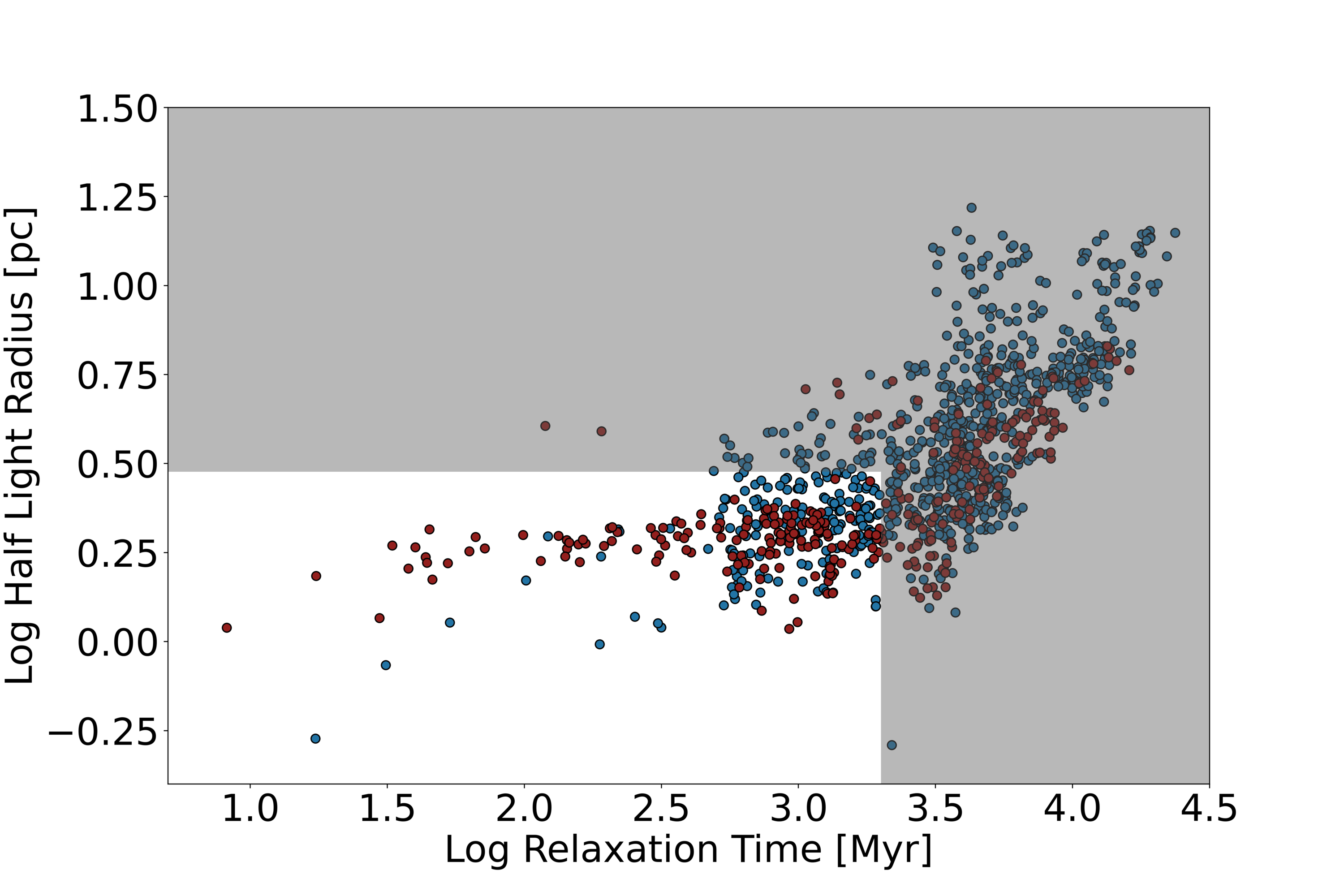} & \includegraphics[width=0.45\textwidth]{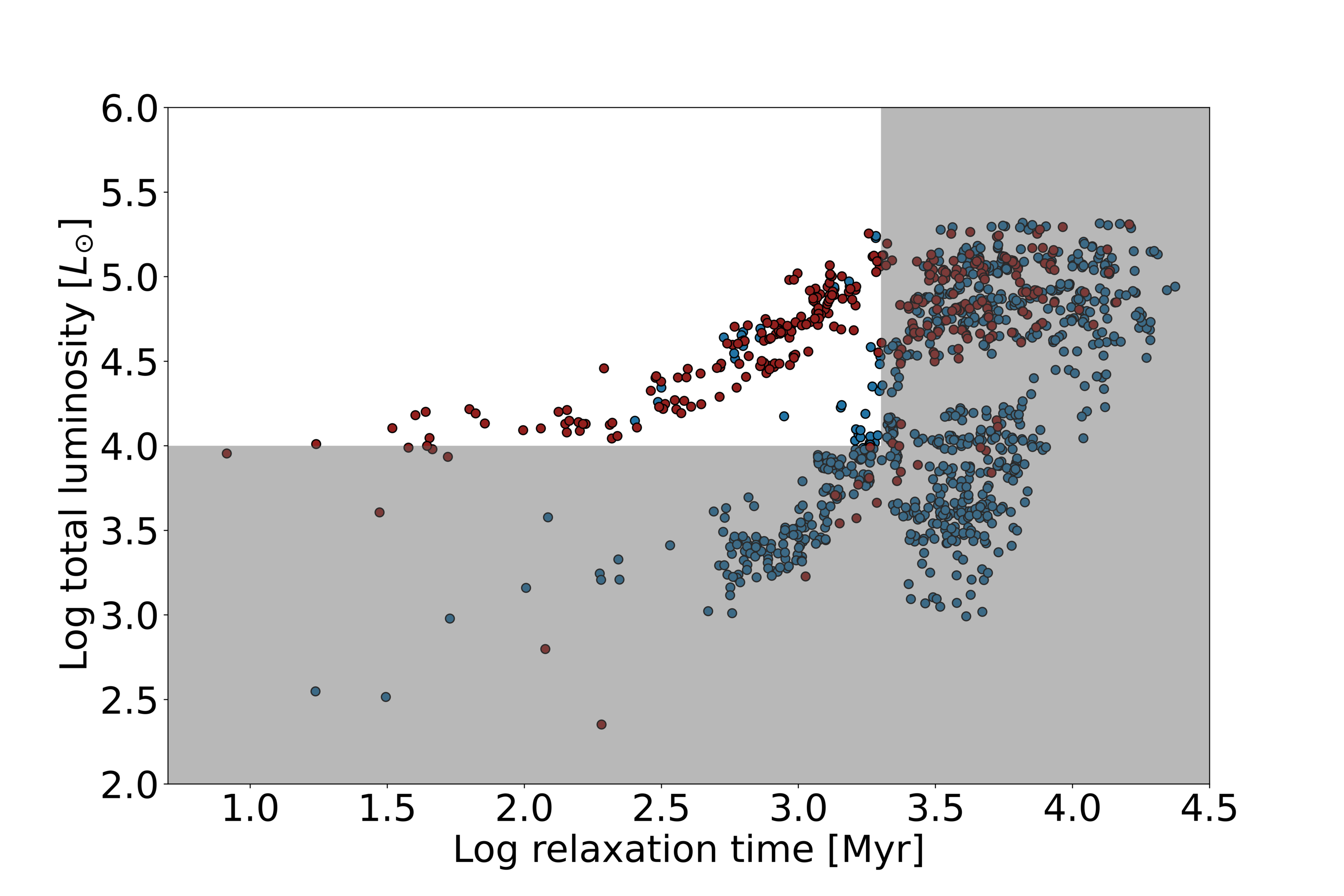} \\
    \includegraphics[width=0.45\textwidth]{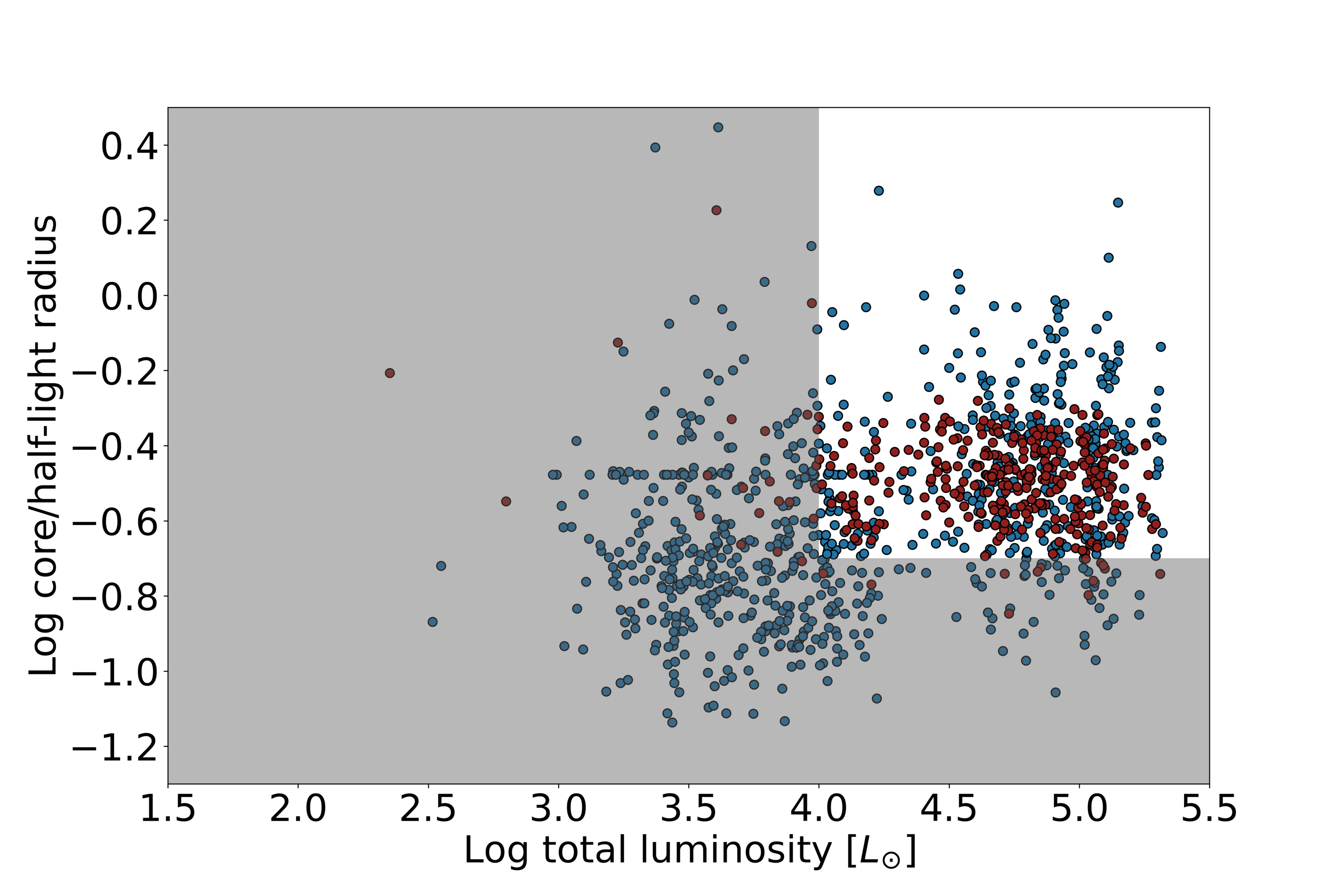} & \includegraphics[width=0.45\textwidth]{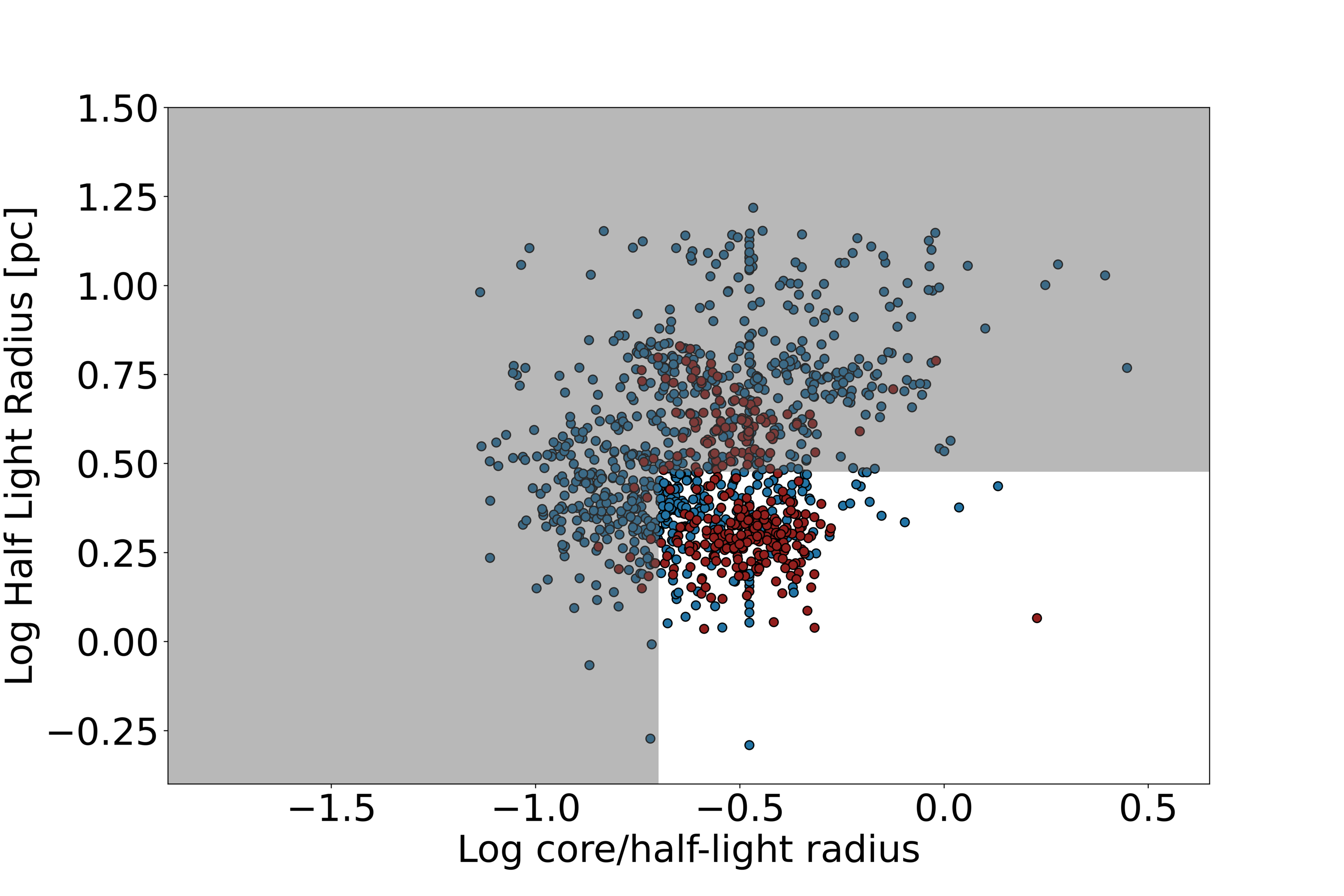} \\
    \end{tabular}
    \caption{The rules found by CORELS, displayed on the planes corresponding each to the relevant couple of features. The white area corresponds to the region that respects the rule. The red points correspond to simulated GCs in the training set that host an IMBH and the blue points correspond to non-hosts.}
    \label{fig:CORELS_TVL_OHLR}
\end{figure}

The rules found by CORELS are reported in Tab.~\ref{rules_found_by_corels} \REFEREE{and shown graphically (in projection on relevant feature planes) in Fig.~\ref{fig:CORELS_TVL_OHLR}.}
They were learned on one of the cross-validation folds each. Remarkably, an identical rule is learned on four folds out of five. The rule learned on the remaining fold is still quite similar to it. All rules stipulate that an IMBH candidate must have an inflated core, in particular requiring that $R_c/R_h > 0.2$. This is immediately interpretable dynamically: a large core is produced by dynamical heating due to a binary containing the IMBH. The IMBH has a high probability of ending up in a binary system in the first place, because its large mass ensures that exchange interactions are favoured in IMBH-binary encounters. In addition, mass segregation makes it likely that the secondary is a stellar-mass black hole. The resulting binary system hardens through interactions with the other objects in the core, releasing energy that ultimately results in a bigger core.

All rules also demand that the total luminosity of a potential candidate exceeds $10^4$~$L_\odot$. This is a very low bar for GCs, whose typical luminosity is about one order of magnitude larger. This rule thus filters out GCs that are likely too small to be able to assemble an IMBH, either because they do not have a high enough density to trigger runaway mergers or because their escape velocity is too small to confine dark remnants that may eventually merge to form an IMBH.

High density is also a criterion for all rules, but it is enforced in one fold by a direct cutoff in central surface brightness, and in the other four by a condition on the half mass radius, i.e. $R_{h} < 3.0$~pc. As shown in Fig.~\ref{fig:CORELS_TVL_OHLR} this combines with the luminosity constraint common to all rules to select high density GCs.

Finally, all rules require that the relaxation time be short, but the cutoffs differ, being set to either $2$ Gyr (in four folds out of five) or to $5$ Gyr (in the remaining one). This is likely the reason for the higher recall in the corresponding fold, as shown in Fig.~\ref{fig:prc}. A short initial relaxation time corresponds to higher interaction frequency between stars, which results in faster mass segregation and makes both runaway mergers and stellar-mass black hole coalescences more likely. Relaxation times observed at $12$ Gyr are essentially a proxy for initial relaxation times, which are unobservable \citep[][]{2010ApJ...708.1598T}. Additionally, the swelling of the core due to an IMBH binary is observable only if the system is relaxed, because the associated dynamical heating requires time to take place. Thus a swollen core in the absence of a short relaxation time is an unreliable indicator of IMBH presence.


\subsection{Rejecting out of distribution points}
\label{KDE}
\begin{figure}
    \centering
    \begin{tabular}{l}
    \hspace{0.2in} \includegraphics[width=0.85\linewidth]{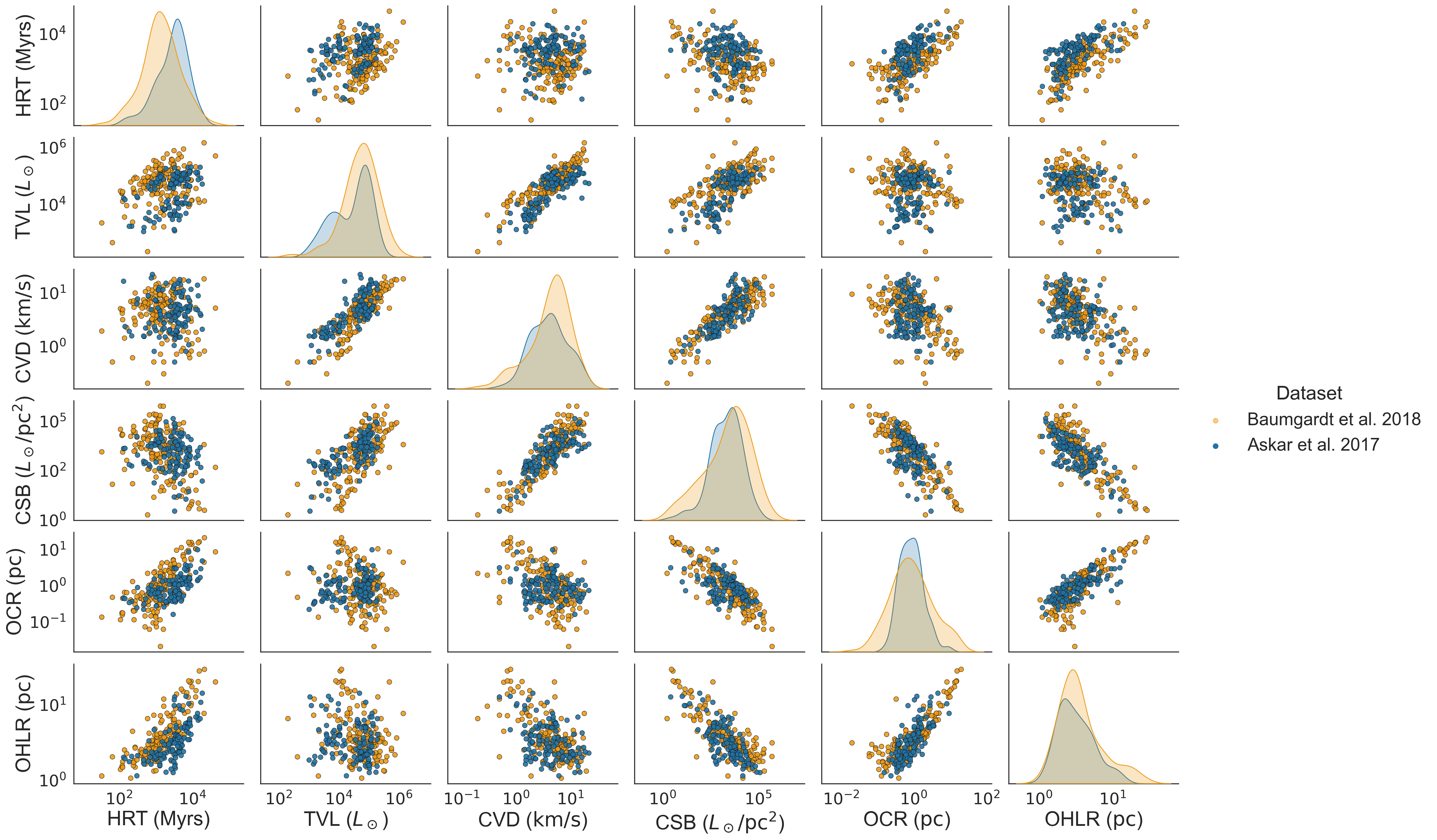} \\    \includegraphics[width=0.8\linewidth]{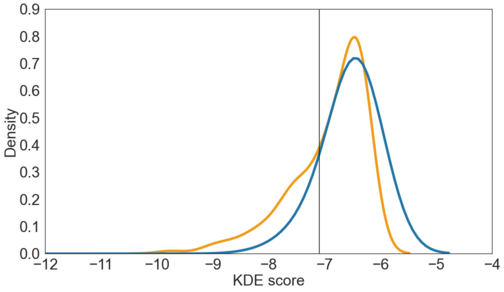} \hspace{0.2in}\\
    \end{tabular}
    \caption{Upper panel: pair plot showing test data (a representative sample of the simulations, shown in blue) versus real GC data (orange). In order, the axis are: Half relaxation time (HRT), total Luminosity (TVL),   central velocity dispersion (CVD),  central surface brightness (CSB), core radius (OCR), and  half-light radius (OHLR). Lower panel: distribution of the KDE score for test data (orange) and actual GC data (blue). The vertical line is the cutoff adopted for inclusion in the final sample of candidates, which corresponds to the first decile of the KDE score for test data. \label{ood}}
\end{figure}

There is a potential mismatch between our simulated training data and actual GCs, which can lead to biased predictions. 
As we can see in Fig.~\ref{ood}, the data on which our models were tested and the real clusters are not perfectly matched.
It is thus crucial to assess how far out-of-distribution the GCs we cast predictions for are with respect to the training data. A multitude of out-of-distribution detection approaches have been discussed in ML literature \citep[see][for a recent review]{2021arXiv211011334Y}. In our case, we stuck to a simple density-based approach, which is generally recognized as effective in low-dimensional feature spaces. For this reason, we utilized KDE.
KDE is a non-parametric technique for estimating the probability density function (PDF) of a random variable from independent, identically distributed samples. The basic idea behind KDE is to place a smooth, symmetric kernel function at each data point, and then sum these kernel functions to estimate the underlying PDF. While other approaches to density estimation \REFEREE{exist}, such as e.g. Gaussian mixture models \citep[][]{mclachlan1988mixture}, mixture density networks \citep[][]{bishop1994mixture}, and autoregressive flows \citep[][]{huang2018neural}, KDE is a simple and time tested method that works well in low dimension. We used a simple Gaussian kernel KDE relying on the scikit-learn package \citep[][]{scikit-learn}.  We fitted the density estimation on the training set and evaluated on both the test set and the real data set. The distributions of the KDE score are shown in Fig.~\ref{ood}. Low KDE score corresponds to weird -i.e. unlikely, out of distribution- data points with respect to the training set. We calculated the KDE scores for the points in our test set (randomly extracted from the simulation data set) and used its first decile as a cutoff to reject the out-of-distribution data points among the actual GCs. This way we remove from the pool of IMBH host candidates those whose prediction would require us to trust our models in regions where the data on which they were trained is scarce. 

\section{Results}
\REFEREE{In this study, we compared a baseline black-box method (XGBoost) to a fully interpretable alternative method (CORELS) for classifying GCs}. \REFEREE{We obtained a comparison between the two methods in terms of performance given the difference in interpretability. To this end, we} utilized \REFEREE{both methods} on the simulated GCs (explained in Section~2) and on real GCs \REFEREE{as well}. \REFEREE{Figure~\ref{fig:prc} shows the precision-recall curve of this classifier} on our test set. Additionally, we applied CORELS, to the same data set, obtaining a precision (recall) of 0.89 (0.46). \REFEREE{Since CORELS returns a hard classification, unlike XGBoost, we did not obtain a precision-recall curve for CORELS, but only single points in the precision-recall plane.} The rules comprising the \REFEREE{rule list learned by CORELS} are visualized in Figure 4.

\begin{figure}
    \centering
    \includegraphics[width=0.9\textwidth]{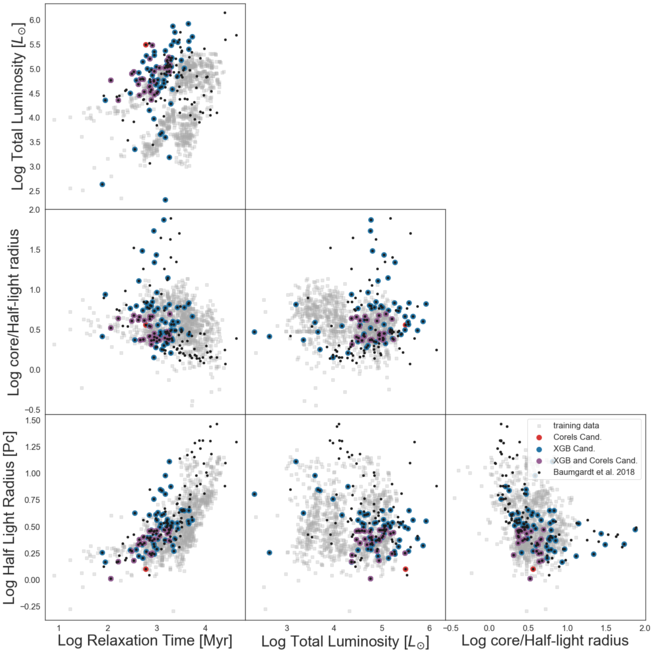}
    \caption{Candidate selection shown in the space of relaxation time, total luminosity, log half-light radius, and log core over half-light radius. The gray squares represent the training data set, the black dots the data from actual GCs compiled by \cite{2018MNRAS.478.1520B}. GCs selected as IMBH host candidates only by XGBoost (CORELS) are shown in blue (red), and candidates selected by both models are shown in purple. 
    \label{fig:candidates_hrt_vs_tvl_hm}}
\end{figure}

\begin{figure}
    \centering
    \includegraphics[width=0.5\linewidth]{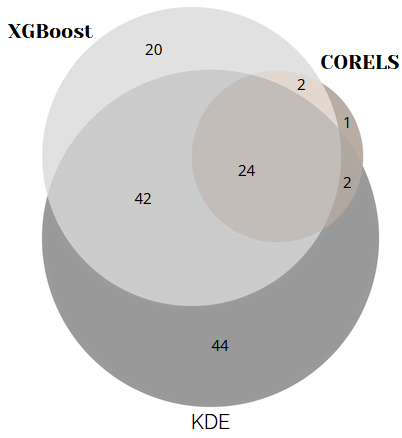}
    \caption{Euler-Venn diagram summarizing the inclusion rules for GCs in the final candidate list. The dark gray set includes all GCs who fall in-distribution with respect to the simulations according to the criterion based on KDE we discussed in the text (112 in total); the light gray set includes the XGBoost candidates (88 in total); and the brown set the CORELS candidates (29 in total). The clusters in the mutual intersection of all three sets are included in the final candidate list (24 in total).}
    \label{fig:euler_venn_diagram}
\end{figure}

Furthermore, we used a \REFEREE{KDE}  approach to measure how much the real data deviated from the test data distribution. We flagged real GCs as out of distribution if they were below the first decile of the KDE score. 
Lastly, we present a table of GCs that are predicted to be IMBH hosts both by CORELS and by XGBoost and also are in-distribution based on our KDE criterion. This is shown in Tab.~\ref{tab:candidates} alongside with the upper limits on IMBH mass obtained for each candidate by T18 (if any are present) and with any reference we found presenting evidence of IMBH presence or absence in the candidate. Candidate selection for this table is summarized by Figs.~\ref{fig:candidates_hrt_vs_tvl_hm} \REFEREE{and  ~\ref{fig:euler_venn_diagram}.}

\begin{table}
\centering
    \caption{Final table of IMBH candidate hosts. The first column reports the cluster name, columns $3$-$8$ report the (rounded, log) values of the relevant features. Columns $9$-$11$ show the classification prediction by XGBoost, CORELS and the in-distribution criterion based on KDE and column $12$ shows the confidence of the classification by XGBoost, based on which the rows are ordered in decreasing order. Column $13$ reports the IMBH upper mass limit based on radio observations by T18. The last column lists any previous claims or other relevant literature discussing IMBHs in the cluster. The full table, including non-hosts, is presented in Appendix \ref{bigtable}.}

\tiny
\begin{tabular}{llllllllllllll}
1 & 2 & 3 & 4 & 5 & 6 & 7 & 8 & 9 & 10 & 11 & 12 & 13 & 14 \\
\hline
 Cluster ID   &  Other ID & HRT &   TVL &   CVD &   CSB &   OCR &   OHLR & CORELS   & XGB  & KDE   &  Pred. proba & T18 upper limit & References \\
\hline
 NGC 6569  &&   3.06 &  5.02 &  0.86 &  3.68 &  0.04 &   0.41 & True          & True       & True     &       0.9999 && \\
 Pal 6     &&   2.67 &  4.79 &  0.72 &  3.64 & -0.02 &   0.35 & True          & True       & True     &       0.9998 && \\
 NGC 6638  &&   2.49 &  4.9  &  0.84 &  4.21 & -0.34 &   0.27 & True          & True       & True     &       0.9998 && \\
 NGC 6333  & M9 &  3.22 &  5.18 &  0.92 &  3.9  & -0.04 &   0.44 & True          & True       & True     &       0.9996 &$1030 M_\odot$& \cite{2007MNRAS.381..103M}\\
 NGC 6712  &&   2.73 &  4.78 &  0.71 &  3.54 &  0.03 &   0.4  & True          & True       & True     &       0.9988 &$1150 M_\odot$& \cite{2007MNRAS.381..103M}\\
   & &    &   &   &   &   &    &          &        &      &        & & \cite{2013AA...558A.117L} \\
 NGC 6254  & M10&   3.25 &  5.04 &  0.79 &  3.65 &  0.02 &   0.47 & True          & True       & True     &       0.9985 &$740 M_\odot$& \cite{2010ApJ...713..194B}\\
   & & & & & & & & & & & &&
\cite{2013ApJ...768...26U}\\
 FSR 1735  &&  2.63 &  4.69 &  0.67 &  3.83 & -0.29 &   0.33 & True          & True       & True     &       0.9969 && \\
 NGC 2298  &&   2.65 &  4.47 &  0.59 &  3.55 & -0.28 &   0.38 & True          & True       & True     &       0.9955 && \cite{2009ApJ...699.1511P} \\
 NGC 5986  &&   3.25 &  5.24 &  0.91 &  3.83 &  0.07 &   0.44 & True          & True       & True     &       0.9954 && \\
 NGC 6218  & M12 &  2.94 &  4.75 &  0.68 &  3.34 &  0.08 &   0.45 & True          & True       & True     &       0.9947 &$800 M_\odot$& \cite{2016MmSAI..87..614S} \\
 NGC 6316  &&  3.27 &  5.09 &  0.91 &  4.04 & -0.22 &   0.47 & True          & True       & True     &       0.9913 && \\
 VVV-CL001 &&  2.75 &  4.54 &  0.81 &  3.65 & -0.27 &   0.34 & True          & True       & True     &       0.9889 && \\
 Ton 2     &&  3.01 &  4.51 &  0.53 &  3.04 &  0.06 &   0.46 & True          & True       & True     &       0.9889 && \\
 NGC 6352  &&  2.96 &  4.47 &  0.54 &  3.06 &  0.05 &   0.46 & True          & True       & True     &       0.9871 && \\
 NGC 6779  &M56&   3.17 &  5.03 &  0.78 &  3.59 &  0.07 &   0.47 & True          & True       & True     &       0.9845 && \\
 NGC 6171  &M107&  2.87 &  4.53 &  0.61 &  3.32 & -0.08 &   0.46 & True          & True       & True     &       0.9843 &$990 M_\odot$& \\
 NGC 6553  &&   3.26 &  4.95 &  0.88 &  3.67 & -0.06 &   0.37 & True          & True       & True     &       0.9837 &&\cite{2016MNRAS.460.2025K} \\
 NGC 6934  &&   3.26 &  4.95 &  0.68 &  3.46 &  0.09 &   0.47 & True          & True       & True     &       0.9815 && \\
 NGC 6637  & M69&  2.95 &  4.95 &  0.79 &  3.76 & -0.07 &   0.38 & True          & True       & True     &       0.9724 && \cite{2007MNRAS.381..103M}\\
 UKS 1     &&  2.85 &  4.63 &  0.62 &  3.29 &  0.04 &   0.46 & True          & True       & True     &       0.9518 && \\
 Ter 1     &&  2.54 &  4.94 &  0.91 &  4.44 & -0.49 &   0.18 & True          & True       & True     &       0.9502 && \\
 NGC 6342  &&  2.22 &  4.36 &  0.64 &  3.85 & -0.47 &   0.17 & True          & True       & True     &       0.9288 && \\
 HP 1      &&  2.89 &  4.58 &  0.72 &  3.18 &  0.1  &   0.46 & True          & True       & True     &       0.8894 && \\
 NGC 6401  &&  2.86 &  4.65 &  0.81 &  3.73 & -0.28 &   0.39 & True          & True       & True     &       0.5381 && \\
\hline
\end{tabular}
    \label{tab:candidates}
\end{table}


\begin{table}
\tiny
\caption{Anchors explaining XGBoost decision for real GCs. The first column reports the GC name, the second column the anchor as a list of conditions (each feature has been given a different colour for ease of comparison), the third column reports the anchor's precision and the fourth column its coverage.\label{ancore}}

\begin{tabular}{llll}
\hline
 Cluster   & Anchor                         &   Precision &   Coverage \\
\hline
 NGC 2298  & ['\textcolor{red}{HRT} $\le$ 741.31', '\textcolor{blue}{CSB} \ensuremath{>} 458.30', {'0.39 \ensuremath{<} \textcolor{cyan}{OCR} $\le$ 1.05'}, '\textcolor{purple}{OHLR} $\le$ 3.15', '\textcolor{green}{TVL} \ensuremath{>} 26722.22', '\textcolor{orange}{CVD} $\le$ 5.10']                     &    0.992 & 0.019  \\
 NGC 5986  & ['\textcolor{green}{TVL} \ensuremath{>} 139719.63', '\textcolor{purple}{OHLR} $\le$ 3.15', {'\textcolor{cyan}{OCR} \ensuremath{>} 1.05'}]                                                                            &    1.000        & 0.006 \\
 NGC 6171  & ['\textcolor{red}{HRT} $\le$ 741.31', '\textcolor{blue}{CSB} \ensuremath{>} 458.30', {'\textcolor{cyan}{OCR} \ensuremath{>} 0.39'}, '2.37 \ensuremath{<} \textcolor{purple}{OHLR} $\le$ 3.15', '\textcolor{orange}{CVD} $\le$ 5.10', '\textcolor{green}{TVL} \ensuremath{>} 26722.22']                      &    1.000        & 0.025  \\
  NGC 6218  & ['\textcolor{red}{HRT} $\le$ 1513.56', '\textcolor{blue}{CSB} \ensuremath{>} 458.30', {'\textcolor{cyan}{OCR} \ensuremath{>} 1.05'}, '\textcolor{purple}{OHLR} $\le$ 3.15']                                                             &    0.990 & 0.056  \\
  FSR 1735  & ['\textcolor{red}{HRT} $\le$ 741.31', '\textcolor{blue}{CSB} \ensuremath{>} 4371.87', {'0.39 \ensuremath{<} \textcolor{cyan}{OCR} $\le$ 1.05'}, '\textcolor{purple}{OHLR} $\le$ 2.37', '\textcolor{green}{TVL} \ensuremath{>} 26722.22', '\textcolor{orange}{CVD} $\le$ 5.10']                    &    0.995 & 0.006 \\
  NGC 6254  & ['\textcolor{green}{TVL} \ensuremath{>} 61963.19', '\textcolor{purple}{OHLR} $\le$ 3.15', '\textcolor{blue}{CSB} $\le$ 16864.96', '5.10 \ensuremath{<} \textcolor{orange}{CVD} $\le$ 7.60']                                                  &    1.000        & 0.050  \\
  NGC 6316  & ['\textcolor{green}{TVL} \ensuremath{>} 61963.19', '\textcolor{purple}{OHLR} $\le$ 3.15', '\textcolor{blue}{CSB} $\le$ 16864.96', '\textcolor{red}{HRT} \ensuremath{>} 1513.56']                                                       &    0.984 & 0.068   \\
  NGC 6333  & ['\textcolor{green}{TVL} \ensuremath{>} 139719.63', '\textcolor{purple}{OHLR} $\le$ 3.15', '\textcolor{blue}{CSB} $\le$ 16864.96', '\textcolor{red}{HRT} \ensuremath{>} 1513.56']                                                      &    1.000        & 0.037  \\
  NGC 6342  & ['\textcolor{red}{HRT} $\le$ 741.31', '4371.87 \ensuremath{<} \textcolor{blue}{CSB} $\le$ 16864.96', '\textcolor{orange}{CVD} \ensuremath{>} 3.10', {'\textcolor{cyan}{OCR} $\le$ 2.15'}]                                                  &    0.870 & 0.074  \\
  NGC 6352  & ['\textcolor{red}{HRT} $\le$ 1513.56', '0.39 \ensuremath{<} \textcolor{cyan}{OCR} $\le$ 2.15', '\textcolor{green}{TVL} \ensuremath{>} 26722.22', '\textcolor{purple}{OHLR} $\le$ 3.15', '\textcolor{orange}{CVD} $\le$ 5.10']                                    &    1.000        & 0.056  \\
 HP 1      & ['\textcolor{red}{HRT} $\le$ 1513.56', '\textcolor{orange}{CVD} \ensuremath{>} 5.10', '0.39 \ensuremath{<} \textcolor{cyan}{OCR} $\le$ 2.15', '2.37 \ensuremath{<} \textcolor{purple}{OHLR} $\le$ 4.54', '\textcolor{blue}{CSB} \ensuremath{>} 458.30']                                &    0.970 & 0.056  \\
 Ter 1     & ['\textcolor{red}{HRT} $\le$ 741.31', '\textcolor{orange}{CVD} \ensuremath{>} 7.60', '61963.19 \ensuremath{<} \textcolor{green}{TVL} $\le$ 139719.63', '\textcolor{cyan}{OCR} $\le$ 1.05', '\textcolor{purple}{OHLR} $\le$ 3.15', '\textcolor{blue}{CSB} \ensuremath{>} 16864.96']              &    0.936 & 0.019  \\
 Ton 2     & ['\textcolor{red}{HRT} $\le$ 1513.56', {'\textcolor{cyan}{OCR} \ensuremath{>} 1.05'}, '\textcolor{purple}{OHLR} $\le$ 3.15']                                                                             &    0.984  & 0.056  \\
 NGC 6401  & ['\textcolor{red}{HRT} $\le$ 741.31', '5.10 \ensuremath{<} \textcolor{orange}{CVD} $\le$ 7.60', '0.39 \ensuremath{<} \textcolor{cyan}{OCR} $\le$ 1.05', '\textcolor{purple}{OHLR} $\le$ 4.54', '4371.87 \ensuremath{<} \textcolor{blue}{CSB} $\le$ 16864.96', '\textcolor{green}{TVL} \ensuremath{>} 26722.22'] &    0.969 & 0.019  \\
 Pal 6     & ['\textcolor{red}{HRT} $\le$ 741.31', '5.10 \ensuremath{<} \textcolor{orange}{CVD} $\le$ 7.60', '0.39 \ensuremath{<} \textcolor{cyan}{OCR} $\le$ 1.05', '\textcolor{purple}{OHLR} $\le$ 2.37', '\textcolor{blue}{CSB} \ensuremath{>} 458.30', '\textcolor{green}{TVL} \ensuremath{>} 26722.22']              &    0.982 & 0.019  \\
 UKS 1     & ['\textcolor{red}{HRT} $\le$ 741.31', '3.10 \ensuremath{<} \textcolor{orange}{CVD} $\le$ 7.60', '0.39 \ensuremath{<} \textcolor{cyan}{OCR} $\le$ 2.15', '\textcolor{green}{TVL} \ensuremath{>} 26722.22', '\textcolor{blue}{CSB} \ensuremath{>} 458.30']                              &    0.945 & 0.062  \\
 VVV-CL001 & ['\textcolor{red}{HRT} $\le$ 741.31', '5.10 \ensuremath{<} \textcolor{orange}{CVD} $\le$ 7.60', '\textcolor{cyan}{OCR} \ensuremath{>} 0.39', '\textcolor{blue}{CSB} \ensuremath{>} 4371.87']                                                       &    0.981 & 0.025  \\
 NGC 6553  & ['\textcolor{green}{TVL} \ensuremath{>} 61963.19', '\textcolor{purple}{OHLR} $\le$ 2.37', '\textcolor{blue}{CSB} $\le$ 16864.96']                                                                        &    1.000        & 0.012  \\
 NGC 6569  & ['\textcolor{red}{HRT} $\le$ 1513.56', '\textcolor{orange}{CVD} \ensuremath{>} 5.10', '\textcolor{purple}{OHLR} \ensuremath{>} 2.37', '\textcolor{blue}{CSB} \ensuremath{>} 4371.87', {'\textcolor{cyan}{OCR} \ensuremath{>} 1.05'}]                                               &    0.992 & 0.006 \\
 NGC 6637  & ['\textcolor{red}{HRT} $\le$ 1513.56', '5.10 \ensuremath{<} \textcolor{orange}{CVD} $\le$ 7.60', '\textcolor{green}{TVL} \ensuremath{>} 61963.19', '\textcolor{blue}{CSB} \ensuremath{>} 4371.87', '\textcolor{purple}{OHLR} \ensuremath{>} 2.37']                                   &    0.993 & 0.031  \\
 NGC 6638  & ['\textcolor{red}{HRT} $\le$ 741.31', '\textcolor{orange}{CVD} \ensuremath{>} 5.10', '\textcolor{cyan}{OCR} \ensuremath{>} 0.39', '\textcolor{purple}{OHLR} $\le$ 2.37']                                                                &    0.983 & 0.025  \\
 NGC 6712  & ['\textcolor{red}{HRT} $\le$ 741.31', '3.10 \ensuremath{<} \textcolor{orange}{CVD} $\le$ 5.10', '0.39 \ensuremath{<} \textcolor{cyan}{OCR} $\le$ 2.15', '\textcolor{purple}{OHLR} $\le$ 3.15', '\textcolor{green}{TVL} \ensuremath{>} 26722.22']                              &    1.000        & 0.031  \\
 NGC 6779  & ['\textcolor{red}{HRT} $\le$ 1513.56', '5.10 \ensuremath{<} \textcolor{orange}{CVD} $\le$ 7.60', '\textcolor{cyan}{OCR} \ensuremath{>} 0.39', '\textcolor{purple}{OHLR} $\le$ 3.15', '\textcolor{green}{TVL} \ensuremath{>} 61963.19']                                     &    1.000        & 0.037  \\
 NGC 6934  & ['\textcolor{green}{TVL} \ensuremath{>} 61963.19', '\textcolor{purple}{OHLR} $\le$ 3.15', {'\textcolor{cyan}{OCR} \ensuremath{>} 1.05'}]                                                                             &    1.000        & 0.031  \\
\hline
\end{tabular}
\end{table}

\subsection{Physical interpretation of the CORELS rule list and of the anchors}
To be classified as an IMBH host by CORELS, a GC has to meet the following conditions: $T_r < 2$ Gyr, $R_{h} < 3.0$ pc, $L > 10^4$ $M_\odot$, and $R_c/R_h > 0.2$. The first rule selects dynamically relaxed GCs. The second and third rules, in combination, correspond to a condition on cluster density. The last rule selects GCs whose core is inflated with respect to the overall size of the cluster. All of these rules can be understood in terms of IMBH formation physics: mass segregation (which brings heavy stars and compact remnants to the cluster centre where they can merge to give rise to an IMBH)  happens on a timescale proportional to the two-body relaxation time. The typical age of a GC is of the order of $\approx 12$ Gyr, so clusters with relaxation times of 2 Gyr or shorter had ample time to undergo mass segregation. 

High density (captured by the requirement to have a small radius for a minimum mass) is instrumental in increasing the frequency of close encounters that can result in IMBH formation. Finally, a dynamically puffed-up core was identified as a sign of IMBH presence early on \citep[][]{2005ApJ...620..238B, 2007MNRAS.379...93H, 2007PASJ...59L..11H}, as IMBHs tend to undergo dynamical exchange reactions with stellar-mass black hole (BH) binaries, forming an IMBH-BH binary that is confined to the core by its heavy mass and releases energy in the core through close encounters, heating it up. Additionally, close interactions with the IMBH can also result in stars merging with it, or they can also get ejected out of the central part of the cluster due to strong encounters. These effects contribute in depleting stars from the central part of the cluster.

There is some overlap between the CORELS rules and the anchors listed in Tab.~\ref{ancore}, suggesting that the latter may also have a similar physical interpretation. However, the anchors explaining the behavior of XGBoost often include a large number of features (even all of them at times) and a disparate array of cutoffs, making a straightforward interpretation harder. These are still a potentially useful tool for investigating a specific candidate in the process of planning a follow-up observation.

\section{Summary}
In the recent years, we witnessed an explosion of ML tools in astronomy.
Still, most contributions avoid or only briefly touch upon the issue of interpretability and explainability of the models they adopt.
In this paper, we distinguish between the two approaches, with interpretability meaning that a model is natively simple enough that a human can understand its inner workings and explainability meaning that a complex or otherwise opaque model is explained post-hoc in terms that a human can understand.
This is a terminological distinction that has not yet been received by the astronomical community, and the trickle of works that address this issue at some level of depth often conflates the two, using interpretability as an umbrella term.

With this in mind, in the present work, we have applied both kinds of techniques, comparing a natively interpretable model -CORELS- with a black-box-plus-explanation model (XGBoost explained using the local model agnostic explanation rules known as anchors) on the problem of IMBH detection in GCs. We trained and validated both models on features measured on GC simulations, and deployed the models in prediction on actual GCs. The features we selected are standard, easily measurable characteristics of GCs and are in fact readily available in multiple catalogues. The meaning of each feature is clearly understood in theoretical terms by astronomers.

We produced a list of GCs that are likely to contain an IMBH according to both models, presenting for each one the anchor rule, the confidence score from XGBoost, and a measure of how far out of distribution each actual GC is with respect to the training data. We compare the rule list learned from the intepretable CORELS model to the anchor rules and conclude that there is a qualitative resemblance but the latter are usually more complex. Moreover, the rule list learned by CORELS has an immediate physical interpretation, some aspects of which are shared with a subset of the anchor rules.

We evaluated the performance of both CORELS and XGBoost on a test set comprised of simulations not seen in training. We find that the CORELS model we learned, based on a rule list comprised of four rules, achieves $90\%$ precision at $\approx 50\%$ recall, whereas the precision achieved by XGBoost at the same recall is nominally perfect ($100\%$). However, this comes at the cost of relying on a black box, which is only moderately mitigated by post-hoc explanations.

Despite the simulator's best efforts, our training data and actual GCs are characterized by distributions that do not perfectly overlap in feature space. This raises the issue of whether we can trust the predictions of models trained on simulations and deployed on observational data. In this context, understanding the inner workings of a model provides a crucial advantage: we can decide whether to trust it based on how its behavior matches our subject-matter knowledge, i.e. the physics of IMBH formation. Still, as an additional safeguard, we removed from our final candidate list GCs that were too far out of distribution with respect to the training data. This was achieved by estimating the density of training data in feature space with KDE and placing a relevant cutoff. This procedure also implicitly yields a measure of the realism of the simulations we considered, with respect to the actual observational data, and taken as a whole. The predictions cast by our models with respect to actual GCs cannot be directly tested, since we do not know which GCs host an IMBH. However, they can be used as guidance for further investigation, knowing that they are based on an interpretable model which is not working in extrapolation.\\

\section*{Acknowledgments}
M. P. acknowledges financial support from the European Union’s Horizon 2020 research and innovation program under the Marie Skłodowska-Curie grant agreement No. 896248. This research was made possible by a generous donation from Eric and Wendy Schmidt by recommendation of the Schmidt Futures program. Y.H. acknowledges support from the National Sciences and Engineering Council of Canada Discovery Grant RGPIN-2020-1505 and the Canada Research Chairs Program. MM  acknowledges financial support from the European Research 
Council for the ERC Consolidator grant DEMOBLACK, under contract no. 
770017 and from the German Excellence Strategy via the Heidelberg Cluster of Excellence (EXC 2181 - 390900948) STRUCTURES.
AA acknowledges support for this paper from project No. 2021/43/P/ST9/03167 co-funded by the
Polish National Science Center (NCN) and the European Union Framework Programme for Research
and Innovation Horizon 2020 under the Marie Skłodowska-Curie grant agreement No.
945339. For the purpose of Open Access, the authors have applied for a CC-BY public copyright licence to any Author Accepted Manuscript (AAM) version arising from this
submission. AA also acknowledges support from the Swedish Research Council through the grant 2017-04217 and from NCN through the grant UMO-2021/41/B/ST9/01191.\\

\bibliographystyle{aasjournal}
\bibliography{references}


\appendix
\section{Full list of candidates}
\label{bigtable}

   
\begin{longtable}{lrrrrrrrlllr}
\hline
 Cluster        &   Final score &   HRT &   TVL &   CVD &   CSB &   OCR &   OHLR & CORELS   & XGB   & KDE   &   Pred. prob. \\

\hline

 NGC\_6569       &        0.9999 &  3.06 &  5.02 &  0.86 &  3.68 &  0.04 &   0.41 & True          & True       & True     &       0.9999 \\

 Pal\_6          &        0.9998 &  2.67 &  4.79 &  0.72 &  3.64 & -0.02 &   0.35 & True          & True       & True     &       0.9998 \\
 NGC\_6638       &        0.9998 &  2.49 &  4.9  &  0.84 &  4.21 & -0.34 &   0.27 & True          & True       & True     &       0.9998 \\
 NGC\_6333       &        0.9996 &  3.22 &  5.18 &  0.92 &  3.9  & -0.04 &   0.44 & True          & True       & True     &       0.9996 \\
 NGC\_6712       &        0.9988 &  2.73 &  4.78 &  0.71 &  3.54 &  0.03 &   0.4  & True          & True       & True     &       0.9988 \\
 NGC\_6254       &        0.9985 &  3.25 &  5.04 &  0.79 &  3.65 &  0.02 &   0.47 & True          & True       & True     &       0.9985 \\
 FSR\_1735       &        0.9969 &  2.63 &  4.69 &  0.67 &  3.83 & -0.29 &   0.33 & True          & True       & True     &       0.9969 \\
 NGC\_2298       &        0.9955 &  2.65 &  4.47 &  0.59 &  3.55 & -0.28 &   0.38 & True          & True       & True     &       0.9955 \\
 NGC\_5986       &        0.9954 &  3.25 &  5.24 &  0.91 &  3.83 &  0.07 &   0.44 & True          & True       & True     &       0.9954 \\
 NGC\_6218       &        0.9947 &  2.94 &  4.75 &  0.68 &  3.34 &  0.08 &   0.45 & True          & True       & True     &       0.9947 \\
 NGC\_6316       &        0.9913 &  3.27 &  5.09 &  0.91 &  4.04 & -0.22 &   0.47 & True          & True       & True     &       0.9913 \\
 VVV-CL001       &        0.9889 &  2.75 &  4.54 &  0.81 &  3.65 & -0.27 &   0.34 & True          & True       & True     &       0.9889 \\
 Ton\_2          &        0.9889 &  3.01 &  4.51 &  0.53 &  3.04 &  0.06 &   0.46 & True          & True       & True     &       0.9889 \\
 NGC\_6352       &        0.9871 &  2.96 &  4.47 &  0.54 &  3.06 &  0.05 &   0.46 & True          & True       & True     &       0.9871 \\
 NGC\_6779       &        0.9845 &  3.17 &  5.03 &  0.78 &  3.59 &  0.07 &   0.47 & True          & True       & True     &       0.9845 \\
 NGC\_6171       &        0.9843 &  2.87 &  4.53 &  0.61 &  3.32 & -0.08 &   0.46 & True          & True       & True     &       0.9843 \\
 NGC\_6553       &        0.9837 &  3.26 &  4.95 &  0.88 &  3.67 & -0.06 &   0.37 & True          & True       & True     &       0.9837 \\
 NGC\_6934       &        0.9815 &  3.26 &  4.95 &  0.68 &  3.46 &  0.09 &   0.47 & True          & True       & True     &       0.9815 \\
 NGC\_6637       &        0.9724 &  2.95 &  4.95 &  0.79 &  3.76 & -0.07 &   0.38 & True          & True       & True     &       0.9724 \\
 UKS\_1          &        0.9518 &  2.85 &  4.63 &  0.62 &  3.29 &  0.04 &   0.46 & True          & True       & True     &       0.9518 \\
 Ter\_1          &        0.9502 &  2.54 &  4.94 &  0.91 &  4.44 & -0.49 &   0.18 & True          & True       & True     &       0.9502 \\
 NGC\_6342       &        0.9288 &  2.22 &  4.36 &  0.64 &  3.85 & -0.47 &   0.17 & True          & True       & True     &       0.9288 \\
 HP\_1           &        0.8894 &  2.89 &  4.58 &  0.72 &  3.18 &  0.1  &   0.46 & True          & True       & True     &       0.8894 \\
 NGC\_6401       &        0.5381 &  2.86 &  4.65 &  0.81 &  3.73 & -0.28 &   0.39 & True          & True       & True     &       0.5381 \\
 NGC\_6440       &        0      &  2.78 &  5.5  &  1.16 &  4.96 & -0.46 &   0.1  & True          & False      & False    &       0.1708 \\
 Djor\_2         &        0      &  3.03 &  4.54 &  0.63 &  3.04 &  0.08 &   0.63 & False         & True       & True     &       0.9951 \\
 NGC\_6539       &        0      &  3.22 &  5.04 &  0.77 &  3.46 &  0.18 &   0.57 & False         & True       & True     &       0.9998 \\
 NGC\_6535       &        0      &  2.54 &  3.92 &  0.43 &  3.88 & -1.1  &   0.43 & False         & False      & True     &       0.0253 \\
 NGC\_6388       &        0      &  3.49 &  5.76 &  1.24 &  4.96 & -0.38 &   0.41 & False         & True       & False    &       0.6810  \\
 NGC\_7006       &        0      &  3.46 &  4.93 &  0.61 &  3.19 &  0.2  &   0.63 & False         & True       & True     &       0.6639 \\
 NGC\_6528       &        0      &  2.46 &  4.5  &  0.67 &  3.93 & -0.48 &   0.28 & False         & True       & True     &       0.9979 \\
 Laevens\_3      &        0      &  2.86 &  3.07 & -0.4  &  0.95 &  0.43 &   0.85 & False         & False      & False    &       0.1948 \\
 NGC\_6522       &        0      &  2.85 &  4.94 &  0.91 &  4.19 & -0.38 &   0.39 & False         & False      & True     &       0.0972 \\
 Djor\_1         &        0      &  3.11 &  4.64 &  0.56 &  3.22 & -0.01 &   0.61 & False         & True       & True     &       0.9980  \\
 NGC\_7078       &        0      &  3.47 &  5.6  &  1.12 &  5.85 & -1.4  &   0.31 & False         & False      & False    &       0.0020  \\
 Ter\_10         &        0      &  3.22 &  4.74 &  0.94 &  4.1  & -0.59 &   0.54 & False         & False      & True     &       0.2129 \\
 NGC\_6517       &        0      &  2.64 &  4.92 &  1    &  4.85 & -0.89 &   0.23 & False         & True       & False    &       0.9365 \\
 Ter\_9          &        0      &  2.38 &  4.57 &  0.89 &  4.23 & -0.6  &   0.2  & False         & False      & True     &       0.1856 \\
 NGC\_6402       &        0      &  3.49 &  5.45 &  1    &  3.67 &  0.33 &   0.55 & False         & False      & True     &       0.0021 \\
 NGC\_6397       &        0      &  2.94 &  4.76 &  0.72 &  5.07 & -1.4  &   0.33 & False         & True       & False    &       0.9645 \\
 Pal\_13         &        0      &  3.26 &  3.19 & -0.4  &  1.02 &  0.29 &   1.11 & False         & False      & False    &       0.4264 \\
 NGC\_6496       &        0      &  3.18 &  4.63 &  0.48 &  2.69 &  0.4  &   0.64 & False         & False      & True     &       0.0069 \\
 NGC\_7089       &        0      &  3.44 &  5.55 &  1.04 &  4.34 & -0.11 &   0.48 & False         & True       & True     &       0.9914 \\

 NGC\_6540       &        0      &  3.09 &  4.39 &  0.4  &  4.23 & -1.22 &   0.42 & False         & False      & True     &       0.0212 \\

 \hline
 Cluster        &   Final score &   HRT &   TVL &   CVD &   CSB &   OCR &   OHLR & CORELS   & XGB   & KDE   &   Pred. prob. \\

\hline
 NGC\_6453       &        0      &  3    &  5.05 &  0.86 &  4.94 & -1    &   0.44 & False         & True       & False    &       0.9967 \\
 NGC\_7099       &        0      &  3.28 &  4.87 &  0.75 &  4.8  & -1.22 &   0.4  & False         & False      & False    &       0.0417 \\
 Ter\_6          &        0      &  2.07 &  4.77 &  0.92 &  4.45 & -0.51 &   0.01 & True          & True       & False    &       0.6464 \\
 Pal\_12         &        0      &  3.11 &  3.69 & -0.15 &  1.36 &  0.59 &   0.84 & False         & True       & False    &       0.8911 \\
 NGC\_6441       &        0      &  3.33 &  5.87 &  1.28 &  5    & -0.28 &   0.32 & False         & True       & False    &       0.8584 \\
 Ter\_5          &        0      &  3.35 &  5.57 &  1.19 &  4.64 & -0.29 &   0.27 & False         & True       & True     &       0.9572 \\
 NGC\_6426       &        0      &  3.43 &  4.56 &  0.45 &  2.48 &  0.45 &   0.71 & False         & False      & True     &       0.0035 \\

 NGC\_6838       &        0      &  3.32 &  4.58 &  0.45 &  2.87 &  0.17 &   0.53 & False         & False      & True     &       0.1592 \\
 NGC\_6541       &        0      &  3.29 &  5.22 &  0.93 &  5.07 & -1.05 &   0.36 & False         & False      & False    &       0.0490  \\
 NGC\_6681       &        0      &  2.71 &  4.8  &  0.85 &  5.12 & -1.15 &   0.33 & False         & True       & False    &       0.9957 \\
 Pal\_11         &        0      &  2.94 &  3.98 &  0.08 &  1.75 &  0.61 &   0.76 & False         & True       & False    &       0.9989 \\
 Ter\_8          &        0      &  4.12 &  4.54 &  0.2  &  1.47 &  1.03 &   1.18 & False         & False      & False    &       0.0024 \\
 NGC\_6809       &        0      &  3.55 &  4.97 &  0.69 &  2.95 &  0.46 &   0.66 & False         & False      & True     &       0.0001 \\
 Arp\_2          &        0      &  3.77 &  4.31 &  0.11 &  1.43 &  0.89 &   1.15 & False         & False      & False    &       0.0093 \\
 Pal\_10         &        0      &  3.35 &  4.94 &  0.66 &  3.09 &  0.32 &   0.6  & False         & False      & True     &       0.0040  \\
 Ter\_7          &        0      &  3.69 &  4.05 &  0.08 &  1.57 &  0.63 &   0.8  & False         & False      & False    &       0.0116 \\
 NGC\_6760       &        0      &  3.28 &  5.17 &  0.83 &  3.72 &  0.03 &   0.51 & False         & True       & True     &       0.9649 \\
 NGC\_6752       &        0      &  3.4  &  5.07 &  0.88 &  4.83 & -0.89 &   0.46 & False         & True       & True     &       0.6079 \\
 NGC\_6749       &        0      &  3.51 &  5.18 &  0.71 &  3.47 &  0.16 &   0.66 & False         & True       & True     &       0.9098 \\
 NGC\_6723       &        0      &  3.21 &  4.94 &  0.74 &  3.3  &  0.23 &   0.55 & False         & False      & True     &       0.1026 \\
 NGC\_6717       &        0      &  2.73 &  4.36 &  0.52 &  3.9  & -0.7  &   0.56 & False         & False      & True     &       0.1662 \\
 NGC\_6715       &        0      &  3.65 &  5.93 &  1.28 &  4.93 & -0.27 &   0.55 & False         & True       & False    &       0.9831 \\
 NGC\_6864       &        0      &  2.95 &  5.35 &  1.05 &  4.69 & -0.42 &   0.31 & False         & True       & True     &       0.9711 \\
 RLGC\_2         &        0      &  3.23 &  5.13 &  0.88 &  4.76 & -0.62 &   0.53 & False         & True       & True     &       0.9972 \\
 Pal\_8          &        0      &  3.15 &  4.03 &  0.49 &  2.25 &  0.29 &   0.54 & False         & False      & True     &       0.0025 \\
 NGC\_6544       &        0      &  2.36 &  4.55 &  0.85 &  4.49 & -0.8  &   0.18 & False         & False      & False    &       0.2371 \\
 NGC\_6656       &        0      &  3.52 &  5.37 &  0.95 &  3.99 &  0.02 &   0.5  & False         & True       & True     &       0.9654 \\
 NGC\_6652       &        0      &  2.19 &  4.43 &  0.69 &  4.1  & -0.59 &   0.17 & False         & False      & True     &       0.3098 \\
 NGC\_6642       &        0      &  1.92 &  4.45 &  0.68 &  4.41 & -0.77 &   0.13 & False         & False      & False    &       0.3241 \\
 NGC\_6626       &        0      &  2.72 &  5.15 &  1.05 &  4.67 & -0.54 &   0.2  & False         & True       & True     &       0.9927 \\
 NGC\_6624       &        0      &  2.91 &  4.87 &  0.85 &  4.37 & -0.48 &   0.37 & False         & False      & True     &       0.3324 \\
 NGC\_6584       &        0      &  3.2  &  4.88 &  0.6  &  3.24 &  0.19 &   0.54 & False         & True       & True     &       0.7763 \\
 BH\_261         &        0      &  2.92 &  3.45 &  0.32 &  2.57 & -0.47 &   0.42 & False         & False      & True     &       0.0000  \\
 NGC\_6981       &        0      &  3.15 &  4.71 &  0.49 &  2.89 &  0.32 &   0.62 & False         & False      & True     &       0.1884 \\
 Ter\_12         &        0      &  2.89 &  4.37 &  0.63 &  3.15 & -0.08 &   0.24 & True          & False      & True     &       0.4512 \\
 IC\_1276        &        0      &  3.07 &  4.61 &  0.56 &  3.12 &  0.07 &   0.5  & False         & True       & True     &       0.9385 \\
 NGC\_6558       &        0      &  1.95 &  4.36 &  0.61 &  4.33 & -0.77 &   0.17 & False         & True       & False    &       0.6577 \\
 2MASS-GC02      &        0      &  2.31 &  3.94 &  0.34 &  2.98 & -0.2  &   0.34 & False         & False      & True     &       0.3450  \\
 ESO\_280-SC06   &        0      &  3.08 &  3.66 & -0.1  &  1.57 &  0.38 &   0.85 & False         & True       & False    &       0.5992 \\
 Sagittarius\_II &        0      &  4.23 &  4.1  & -0.22 &  0.47 &  1.31 &   1.46 & False         & False      & False    &       0.0020  \\
 2MASS-GC01      &        0      &  2.81 &  4.29 &  0.43 &  3.28 & -0.3  &   0.55 & False         & False      & True     &       0.0202 \\
 NGC\_104        &        0      &  3.73 &  5.66 &  1.08 &  4.34 & -0.1  &   0.56 & False         & True       & True     &       0.9550  \\
 NGC\_6380       &        0      &  3.2  &  5.06 &  0.91 &  3.68 &  0.05 &   0.5  & False         & True       & True     &       0.9987 \\
 NGC\_4147       &        0      &  2.74 &  4.38 &  0.51 &  3.62 & -0.44 &   0.4  & False         & False      & True     &       0.0800   \\
 Rup\_106        &        0      &  3.58 &  4.39 &  0.2  &  1.9  &  0.71 &   0.88 & False         & False      & True     &       0.0298 \\
 NGC\_4590       &        0      &  3.53 &  4.83 &  0.57 &  2.96 &  0.29 &   0.65 & False         & False      & True     &       0.0003 \\
 BH\_140         &        0      &  3.48 &  4.38 &  0.38 &  2    &  0.68 &   0.79 & False         & False      & True     &       0.0008 \\
 NGC\_4833       &        0      &  3.22 &  5.07 &  0.79 &  3.49 &  0.19 &   0.51 & False         & True       & True     &       0.9994 \\

   \hline
 Cluster        &   Final score &   HRT &   TVL &   CVD &   CSB &   OCR &   OHLR & CORELS   & XGB   & KDE   &   Pred. prob. \\

\hline

 NGC\_5024       &        0      &  3.98 &  5.41 &  0.8  &  3.36 &  0.36 &   0.81 & False         & False      & True     &       0.0005 \\

 NGC\_5053       &        0      &  3.97 &  4.46 &  0.3  &  1.49 &  1.01 &   1.09 & False         & False      & False    &       0.0000      \\
 NGC\_5139       &        0      &  4.4  &  6.15 &  1.25 &  3.82 &  0.63 &   0.88 & False         & False      & False    &       0.0127 \\
 NGC\_5272       &        0      &  3.66 &  5.4  &  0.88 &  3.87 &  0.09 &   0.53 & False         & True       & True     &       0.9363 \\
 NGC\_5286       &        0      &  3.18 &  5.38 &  0.99 &  4.6  & -0.4  &   0.37 & False         & False      & True     &       0.4562 \\
 AM\_4           &        0      &  3.18 &  2.32 & -0.7  &  0.12 &  0.33 &   0.8  & False         & True       & False    &       0.7439 \\
 NGC\_5466       &        0      &  3.81 &  4.62 &  0.28 &  1.93 &  0.81 &   0.98 & False         & False      & False    &       0.0001 \\
 NGC\_5634       &        0      &  3.57 &  5.02 &  0.75 &  3.79 & -0.12 &   0.65 & False         & True       & True     &       0.9919 \\
 NGC\_5694       &        0      &  3.29 &  5.18 &  0.96 &  5.38 & -1.4  &   0.49 & False         & True       & False    &       0.8416 \\
 IC\_4499        &        0      &  4.02 &  4.83 &  0.48 &  2.22 &  0.71 &   1.01 & False         & False      & True     &       0.0000      \\
 NGC\_5824       &        0      &  3.66 &  5.57 &  1.05 &  4.42 & -0.18 &   0.65 & False         & True       & True     &       0.9533 \\
 NGC\_4372       &        0      &  3.68 &  4.98 &  0.65 &  2.74 &  0.58 &   0.76 & False         & False      & True     &       0.0002 \\
 Crater          &        0      &  3.78 &  3.97 & -0.22 &  0.74 &  1.1  &   1.29 & False         & False      & False    &       0.0562 \\
 NGC\_5897       &        0      &  3.82 &  4.85 &  0.54 &  2.34 &  0.74 &   0.88 & False         & False      & True     &       0.0000      \\
 Pal\_4          &        0      &  3.72 &  4.28 & -0.15 &  1.17 &  1.06 &   1.2  & False         & False      & False    &       0.1473 \\
 NGC\_362        &        0      &  3.16 &  5.29 &  0.93 &  4.49 & -0.4  &   0.34 & False         & True       & True     &       0.7864 \\
 Whiting\_1      &        0      &  3.18 &  3.6  & -0.4  &  1.36 &  0.39 &   0.98 & False         & True       & False    &       0.5067 \\
 NGC\_1261       &        0      &  3.3  &  5.05 &  0.76 &  3.79 & -0.06 &   0.51 & False         & True       & True     &       0.8986 \\
 Pal\_1          &        0      &  1.89 &  2.63 & -0.3  &  1.56 & -0.16 &   0.26 & False         & True       & False    &       0.9337 \\
 AM\_1           &        0      &  3.63 &  4.37 & -0.05 &  1.54 &  0.82 &   1.17 & False         & False      & False    &       0.1989 \\
 Eridanus        &        0      &  3.48 &  4.1  & -0.15 &  1.29 &  0.83 &   1.14 & False         & False      & False    &       0.3368 \\
 Pal\_2          &        0      &  3.67 &  5.24 &  0.71 &  3.44 &  0.21 &   0.68 & False         & False      & True     &       0.1936 \\
 NGC\_1851       &        0      &  2.95 &  5.28 &  1.04 &  5.32 & -1.1  &   0.24 & False         & True       & False    &       0.6323 \\
 NGC\_1904       &        0      &  2.81 &  5    &  0.81 &  4.3  & -0.38 &   0.4  & False         & True       & True     &       0.8813 \\
 NGC\_2419       &        0      &  4.63 &  5.69 &  0.76 &  2.66 &  0.91 &   1.29 & False         & False      & False    &       0.0000      \\
 Pyxis           &        0      &  3.9  &  4.13 &  0    &  0.93 &  1.13 &   1.23 & False         & False      & False    &       0.0135 \\
 NGC\_2808       &        0      &  3.37 &  5.76 &  1.15 &  4.62 & -0.1  &   0.39 & False         & True       & False    &       0.8892 \\
 E\_3            &        0      &  2.56 &  3.35 & -0.22 &  1.59 &  0.26 &   0.63 & False         & True       & False    &       0.9953 \\
 Pal\_3          &        0      &  3.95 &  4.11 & -0.1  &  0.8  &  1.16 &   1.31 & False         & False      & False    &       0.0008 \\
 NGC\_3201       &        0      &  3.51 &  4.87 &  0.66 &  3.33 &  0.09 &   0.58 & False         & False      & True     &       0.0182 \\
 Pal\_5          &        0      &  3.87 &  3.91 & -0.22 &  0.57 &  1.17 &   1.31 & False         & False      & False    &       0.0028 \\
 NGC\_5904       &        0      &  3.51 &  5.34 &  0.9  &  3.86 &  0.07 &   0.55 & False         & True       & True     &       0.6028 \\
 Liller\_1       &        0      &  2.85 &  5.53 &  1.34 &  5.33 & -0.72 &   0.05 & False         & False      & False    &       0.1790  \\
 NGC\_6266       &        0      &  2.91 &  5.49 &  1.18 &  4.85 & -0.41 &   0.25 & True          & True       & False    &       0.9919 \\
 NGC\_6284       &        0      &  2.89 &  4.99 &  0.77 &  4.14 & -0.33 &   0.47 & False         & True       & True     &       0.9988 \\
 NGC\_6287       &        0      &  3.01 &  4.71 &  0.8  &  4.21 & -0.68 &   0.29 & False         & True       & True     &       0.7655 \\
 NGC\_6293       &        0      &  3.16 &  4.91 &  0.88 &  4.61 & -0.89 &   0.37 & False         & False      & True     &       0.4187 \\
 NGC\_6304       &        0      &  3.1  &  4.9  &  0.74 &  4.38 & -0.66 &   0.29 & False         & True       & True     &       0.7501 \\
 NGC\_6325       &        0      &  2.24 &  4.45 &  0.72 &  4.05 & -0.51 &   0.23 & False         & False      & True     &       0.3542 \\
 NGC\_6341       &        0      &  3.39 &  5.21 &  0.92 &  4.01 & -0.11 &   0.38 & False         & True       & True     &       0.6153 \\
 NGC\_6355       &        0      &  2.86 &  4.79 &  0.74 &  4.34 & -0.68 &   0.37 & False         & False      & True     &       0.0326 \\
 NGC\_6356       &        0      &  3.72 &  5.34 &  0.94 &  3.64 &  0.17 &   0.59 & False         & True       & True     &       0.5412 \\
 IC\_1257        &        0      &  2.85 &  4.16 &  0.23 &  2.66 &  0.06 &   0.61 & False         & True       & True     &       0.7621 \\
 Ter\_2          &        0      &  3.03 &  4.78 &  0.76 &  3.95 & -0.38 &   0.39 & False         & True       & True     &       0.8764 \\
 Ter\_4          &        0      &  3.43 &  4.85 &  0.74 &  3.67 & -0.24 &   0.54 & False         & False      & True     &       0.4998 \\
 NGC\_288        &        0      &  3.47 &  4.64 &  0.49 &  2.43 &  0.56 &   0.77 & False         & False      & True     &       0.0024 \\
 FSR\_1758       &        0      &  4.41 &  5.6  &  0.76 &  2.63 &  1.03 &   1.1  & False         & False      & False    &       0.0000      \\
 NGC\_6362       &        0      &  3.42 &  4.83 &  0.59 &  2.78 &  0.46 &   0.71 & False         & False      & True     &       0.0010  \\

   \hline
 Cluster        &   Final score &   HRT &   TVL &   CVD &   CSB &   OCR &   OHLR & CORELS   & XGB   & KDE   &   Pred. prob. \\

\hline
 
 NGC\_6366       &        0      &  2.99 &  4.34 &  0.38 &  2.58 &  0.29 &   0.58 & False         & False      & True     &       0.0010  \\

 NGC\_6273       &        0      &  3.3  &  5.49 &  1.08 &  4.19 &  0    &   0.5  & False         & True       & True     &       0.9990  \\
 Pal\_15         &        0      &  4.1  &  4.16 &  0.11 &  0.86 &  1.16 &   1.3  & False         & False      & False    &       0.0003 \\
 NGC\_5927       &        0      &  3.31 &  5.23 &  0.83 &  3.68 &  0.15 &   0.54 & False         & True       & True     &       0.6543 \\
 NGC\_6256       &        0      &  3.15 &  4.77 &  0.73 &  4.9  & -1.4  &   0.47 & False         & True       & False    &       0.8825 \\
 NGC\_5946       &        0      &  2.59 &  4.77 &  0.78 &  4.22 & -0.51 &   0.29 & False         & True       & True     &       0.9940  \\
 FSR\_1716       &        0      &  3.01 &  3.84 &  0.52 &  2.13 &  0.28 &   0.56 & False         & False      & True     &       0.0063 \\
 Pal\_14         &        0      &  4.1  &  4.05 & -0.15 &  0.55 &  1.21 &   1.44 & False         & False      & False    &       0.0006 \\
 Lynga\_7        &        0      &  2.99 &  4.68 &  0.56 &  2.91 &  0.33 &   0.61 & False         & False      & True     &       0.2345 \\
 NGC\_6093       &        0      &  2.86 &  5.21 &  1.04 &  4.53 & -0.39 &   0.25 & True          & False      & True     &       0.4276 \\
 RLGC\_1         &        0      &  3.9  &  5.22 &  0.66 &  3    &  0.45 &   0.9  & False         & False      & True     &       0.0000      \\
 NGC\_6101       &        0      &  4.04 &  4.84 &  0.52 &  2.21 &  0.77 &   0.98 & False         & False      & True     &       0.0000      \\
 NGC\_6121       &        0      &  2.87 &  4.74 &  0.68 &  3.96 & -0.41 &   0.4  & False         & True       & True     &       0.6485 \\
 NGC\_6139       &        0      &  2.79 &  5.27 &  1.06 &  4.87 & -0.57 &   0.27 & False         & True       & False    &       0.8153 \\
 NGC\_6144       &        0      &  3.01 &  4.5  &  0.57 &  2.88 &  0.21 &   0.56 & False         & True       & True     &       0.7837 \\
 Ter\_3          &        0      &  3.17 &  4.36 &  0.34 &  2.32 &  0.47 &   0.67 & False         & True       & True     &       0.5244 \\
 ESO\_452-SC11   &        0      &  2.41 &  3.53 &  0.15 &  2.42 & -0.14 &   0.4  & False         & False      & True     &       0.2605 \\
 NGC\_6205       &        0      &  3.52 &  5.37 &  0.97 &  3.66 &  0.28 &   0.54 & False         & False      & True     &       0.0004 \\
 NGC\_6229       &        0      &  3.21 &  5.17 &  0.88 &  3.8  &  0.02 &   0.49 & False         & True       & True     &       0.9999 \\
 NGC\_6235       &        0      &  3.09 &  4.62 &  0.64 &  3.15 &  0.08 &   0.52 & False         & True       & True     &       0.9363 \\
 NGC\_7492       &        0      &  3.29 &  4.29 &  0.18 &  1.89 &  0.67 &   0.88 & False         & True       & True     &       0.8244 \\
\hline
\multicolumn{12}{p{\textwidth}}{}\\
\multicolumn{12}{p{\textwidth}}{Table A1: \small{Whole sample of star clusters including deemed non-hosts. The ordering of the star clusters in the table is decreasing with `final score' (column 2). This is the XGBoost predicted probability for the star clusters that were classified as hosts by CORELS (column 9) and XGBoost (column 10) and were deemed in-distribution based on our KDE criterion (column 11). We also report cluster features (columns 3--8).}}\\
\end{longtable}

\end{document}